\documentclass[twocolumn]{aastex631}

\usepackage[T1]{fontenc}
\usepackage{ae,aecompl}
\usepackage{natbib}

\usepackage{graphicx}	
\usepackage{amsmath}	
\usepackage{amssymb}	
\usepackage{bm}		
\usepackage{booktabs}
\usepackage{mathtools}
\usepackage{makecell}
\usepackage{bm}
\usepackage{float}
\usepackage[utf8x]{inputenc} 
\usepackage{color,soul}
\usepackage{breqn}

\newcommand{\fesc}{\ensuremath{f_{\rm esc}}}

\newcommand{\woiii}{\ensuremath{W_{\lambda}}(\rm [O~\textsc{iii}])}

\newcommand{\luv}{\ensuremath{L_{\rm UV}}}

\newcommand{\mstar}{\ensuremath{\rm M_{\rm *}}}
\newcommand{\muv}{\ensuremath{\rm M_{\rm UV}}}

\newcommand{\zstar}{\ensuremath{Z_{\rm *}}}
\newcommand{\zsun}{\ensuremath{Z_{\odot}}}

\newcommand{\xiion}{\ensuremath{\xi_{\rm ion}}}
\newcommand{\xiiono}{\ensuremath{\xi_{\rm ion,0}}}
\newcommand{\oh}{\ensuremath{12+\textrm{log(O/H)}}}
\newcommand{\av}{\ensuremath{\textrm{A}_{\rm V}}}
\newcommand{\aha}{\ensuremath{\textrm{A}_{\mathrm{H}\alpha}}}
\newcommand{\lxiion}{\textrm{log}\ensuremath{_{10}(\xi_{\rm ion}/\textrm{erg\:Hz}^{-1})}}
\newcommand{\lxiiono}{\textrm{log}\ensuremath{_{10}(\xi_{\rm ion,0}/\textrm{erg\:Hz}^{-1})}}
\newcommand{\ebmv}{\ensuremath{E(B-V)}}
\newcommand{\ebmvneb}{\ensuremath{E(B-V)_{\rm neb}}}
\newcommand{\ebmvst}{\ensuremath{E(B-V)_{\rm stellar}}}

\newcommand{\aindv}{AURORA\ensuremath{_{\rm indv}}}

\newcommand{\lya}{Ly$\alpha$}
\newcommand{\oiii}{[O~\textsc{iii}]\ensuremath{\lambda\lambda4959,5007}}
\newcommand{\oiiif}{[O~\textsc{iii}]\ensuremath{\lambda5007}}
\newcommand{\oii}{[O~\textsc{ii}]\ensuremath{\lambda\lambda3726,3729}}
\newcommand{\hb}{H\ensuremath{\beta}}
\newcommand{\ha}{H\ensuremath{\alpha}}
\newcommand{\hg}{H\ensuremath{\gamma}}
\newcommand{\hd}{H\ensuremath{\delta}}
\newcommand{\he}{H\ensuremath{\epsilon}}

\newcommand{\nii}{[N~\textsc{ii}]\ensuremath{\lambda6585}}
\newcommand{\oiiia}{[O~\textsc{iii}]\ensuremath{\lambda4363}}
\newcommand{\oiia}{[O~\textsc{ii}]\ensuremath{\lambda\lambda7322,7332}}
\newcommand{\siiia}{[S~\textsc{iii}]\ensuremath{\lambda6314}}
\newcommand{\siia}{[S~\textsc{ii}]\ensuremath{\lambda4070}}

\shorttitle{AURORA: \xiion{}}
\shortauthors{Pahl et al.}

\begin{document}
	
	\title[.]{The AURORA Survey: Ionizing Photon Production Efficiency with Minimal Nebular Dust Attenuation Systematics}
	
	\author[0000-0003-4464-4505]{Anthony J. Pahl}
	\altaffiliation{Carnegie Fellow}
	\affiliation{The Observatories of the Carnegie Institution for Science, 813 Santa Barbara Street, Pasadena, CA 91101, USA}
	
	\author[0000-0003-3509-4855]{Alice Shapley}
	\affiliation{Department of Physics and Astronomy, University of California, Los Angeles, CA 90095, USA}
	
	\author[0000-0001-9687-4973]{Naveen A. Reddy}
	\affiliation{Department of Physics and Astronomy, University of California Riverside, Riverside, CA 92521, USA}
	
	\author[0000-0003-4792-9119]{Ryan Sanders}
	\affiliation{University of Kentucky, 506 Library Drive, Lexington, KY, 40506, USA}
	
	\author[0000-0001-8426-1141]{Michael W. Topping}
	\affiliation{Steward Observatory, University of Arizona, 933 N. Cherry Avenue, Tucson, AZ 85721, USA}

    \author[0000-0002-4153-053X]{Danielle A. Berg}
    \affiliation{Department of Astronomy, The University of Texas at Austin, 2515 Speedway, Stop C1400, Austin, TX 78712, USA}
    
    \author[0000-0002-7622-0208]{Callum T. Donnan}
    \affiliation{NSF's National Optical-Infrared Astronomy Research Laboratory, 950 N. Cherry Ave., Tucson, AZ 85719, USA}

    \author[0000-0002-1404-5950]{James S. Dunlop}
    \affiliation{Institute for Astronomy, University of Edinburgh, Royal Observatory, Edinburgh EH9 3HJ, UK}

    \author[0000-0001-7782-7071]{Richard S. Ellis}
    \affiliation{University College London, Department of Physics \& Astronomy, Gower Street, London WC1E 6BT, UK}
    
    \author[0000-0003-4264-3381]{N. M. F\"orster Schreiber}
	\affiliation{Max-Planck-Institut f\"ur extraterrestrische Physik (MPE), Giessenbachstr.1, D-85748 Garching, Germany}

    \author[0000-0002-3254-9044]{K. Glazebrook}
    \affiliation{Centre for Astrophysics and Supercomputing, Swinburne University of Technology, PO Box 218, Hawthorn, VIC 3122, Australia}

    \author[0000-0003-4368-3326]{Derek J. McLeod}
    \affiliation{Institute for Astronomy, University of Edinburgh, Royal Observatory, Edinburgh EH9 3HJ, UK}
    
    \author[0000-0002-5139-4359]{Max Pettini}
    \affiliation{Institute of Astronomy, University of Cambridge, Madingley Road, Cambridge CB3 0HA, UK}

    \author[0000-0001-7144-7182]{Daniel Schaerer}
    \affiliation{Observatoire de Gen\`eve, Universit\'e de Gen\`eve, Chemin Pegasi 51, 1290 Versoix, Switzerland}
    \affiliation{CNRS, IRAP, 14 Avenue E. Belin, 31400 Toulouse, France}

	\begin{abstract}
		We present ionizing photon production efficiencies (\xiion) for 63 $z=1.5-6.9$ star-forming galaxies using precise nebular dust attenuation corrections from the \textit{JWST}/AURORA survey. A subset of objects within AURORA have individually-determined nebular dust attenuation curves, which vary significantly in shape and normalization, resulting in reduced systematic uncertainty when constraining the total attenuation of \ha{} luminosity, and thus the intrinsic ionizing output within our sample. 
        We find evidence for positive correlations between \xiion{} and redshift, equivalent width of \oiiif{}, and $\textrm{O32}$=\oiiif{}/\oii{}, and negative correlations between \xiion{} and stellar attenuation, UV luminosity (\luv{}), stellar mass, and direct-method metallicity. 
        We test alternate dust prescriptions within this sample, and find that the total attenuation is lower when using the commonly-assumed Galactic extinction curve or when assuming that stellar attenuation is equal to nebular attenuation. We also find that assuming either of these alternate dust prescriptions can change the slope of relationships between \xiion{} and galaxy property, notably inducing a flat trend between \xiion{} and \luv{} within AURORA. While the novel nebular dust curves derived from AURORA spectroscopy reveal obscured ionizing photon production within star-forming galaxies at these redshifts, a more complete understanding of stellar attenuation is required to fully reduce dust systematics on \xiion{} for inclusion in reionization models.
	\end{abstract}
	
	\keywords{Galaxy evolution (594), High-redshift galaxies (734), Interstellar dust (836), Near infrared astronomy (1093), Reionization (1383)}

	\defcitealias{reddyJWSTAURORASurvey2025}{R25}
	\section{Introduction} \label{sec:intro}
	
	Cosmic reionization describes a key phase change in the Universe's history, in which Hydrogen in the cosmic web transitioned from neutral to ionized. To understand this process, one must have a full description of the astrophysical sources of ionizing photons as well as the sinks that absorb them.
	In terms of sinks of H~\textsc{i} in the intergalactic medium (IGM), measurement of the optical depth of free electrons, the mean free path of ionizing photons, and \lya{} damping wings in early galaxies indicate that reionization is well underway at $z\sim7-8$ with an end at $z\sim5.3-6$ \citep[e.g.,][]{planckcollaborationPlanck2015Results2016,beckerMeanFreePath2021,masonConstraints$zsim613$Intergalactic2025}. Despite the growing number of neutral fraction estimates, significant uncertainties and systematic effects mean that the timeline of reionization remains poorly constrained.
	
	The sources of ionizing photons in the epoch of reionization are thought to be primarily massive stars within star-forming galaxies \citep[e.g.][]{rosdahlSPHINXCosmologicalSimulations2018,shenBolometricQuasarLuminosity2020}, although uncertain contributions from AGN have been recently suggested \citep{maiolinoJADESDiversePopulation2024,madauCosmicReionizationJWST2024}.
	To estimate the contribution of stars to the cosmic ionizing emissivity, one can examine three parameters: the total comoving, non-ionizing UV luminosity density $\rho_{\rm UV}$, the ionizing photon production efficiency \xiion{}, and the escape fraction of ionizing photons \fesc{} \citep{robertsonCosmicReionizationEarly2015}. A complete understanding of the production and escape of ionizing photons across a range of galaxy properties is thus required to understand which galaxy populations drive reionization, and produce complete models of the evolution of the neutral fraction of H~\textsc{i}.
	With the advent of the \textit{James Webb Space Telescope} (\textit{JWST}), accessibility of these parameters has grown in fidelity and cosmic volume, with the exception of direct measures of \fesc{}, which require rest-UV observations of the Lyman continuum (LyC) of galaxies at $z\lesssim4$ \citep{vanzellaDetectionIonizingRadiation2012}. Reionization models are particularly sensitive to choices of \xiion{}, defined as the ratio of intrinsic ionizing output per unit UV luminosity density, and its dependence on galaxy properties. In \citet{munozReionizationJWSTPhoton2024a}, the authors outline an ``ionizing photon budget crisis," where early \textit{JWST}-based UV luminosity functions and constraints on \xiion{} along with reasonable assumptions of \fesc{} lead to far too many ionizing photons entering the IGM when compared to independent measurements of the IGM neutral fraction \citep[e.g.][]{mcgreerModelindependentEvidenceFavour2015,masonModelindependentConstraintsHydrogenionizing2019,nakaneLyaEmission132024}. While additional uncertainty exists in choice of \fesc{} and the faint-end cutoff the UVLF, early \xiion{} results from \textit{JWST} were higher than previously-assumed values \citep[$\lxiion=25.3-25.6$,][]{boyettExtremeEmissionLine2024a,saxenaJADESProductionEscape2024,prieto-lyonProductionIonizingPhotons2023} and demonstrated an inverse relation between \xiion{} and UV luminosity \citep[\luv{},][]{simmondsIonizingPropertiesGalaxies2024} contributed to this crisis.
	 
	In more recent analyses of \xiion{} featuring larger, more complete samples and measurements using spectroscopy, median \xiion{} values appeared to be lower than initially found with \textit{JWST} \citep[$\lxiion=25.2-25.5$,][]{simmondsLowmassBurstyGalaxies2024,pahlSpectroscopicAnalysisIonizing2025,llerenaIonizingPhotonProduction2025,papovichGalaxiesEpochReionization2025}. Additionally, updated trends between \xiion{} and \luv{} were found to be flat or slightly positive \citep{simmondsLowmassBurstyGalaxies2024,pahlSpectroscopicAnalysisIonizing2025}. The change in slope between these two parameters significantly reduces the cumulative number of ionizing photons produced over cosmic time, easing the crisis \citep{pahlSpectroscopicAnalysisIonizing2025}. Still, other studies using either large spectroscopic samples with detections of a single Balmer emission line \citep{llerenaIonizingPhotonProduction2025} or with spectrophotometric fits to both \xiion{} and \fesc{} \citep{papovichGalaxiesEpochReionization2025} continue to find that faint galaxies are $\sim2-15\times$ more efficient at producing ionizing photons than their UV bright counterparts. When we additionally consider the uncertain trend between \xiion{} and stellar mass \citep[\mstar, e.g.,][]{begleyEvolutionIIIHv2025,pahlSpectroscopicAnalysisIonizing2025}, the population of galaxies that drive the reionization process still remains under debate.

    A key uncertainty in the determination of \xiion{} is the estimate of dust attenuation, required to constrain the intrinsic efficiency of ionizing photon production. This reddening affects both the estimate of ionizing photons produced, commonly extrapolated from the strength of Balmer recombination lines like \ha{} in spectroscopic studies, and \luv{} (evaluated at $\lambda_{\rm rest}\sim1500$\AA{}) which normalizes the efficiency. 
    Stellar (attenuating \luv{}) and nebular (attenuating \ha{} line strength) dust curves are dependent on dust grain size, absorption and scattering properties, spatial distribution, and column density. These properties may be different depending how similar the sightlines are to massive O stars and less-massive stars contributing to \luv{} \citep[e.g.,][]{wildEmpiricalDeterminationShape2011,priceDirectMeasurementsDust2014}. For star-forming galaxies at high-redshift, fitting the rest-UV stellar continuum is very sensitive to reddening, and the \citet{calzettiDustContentOpacity2000} or Small Magellanic Cloud \citep[SMC,][]{gordonQuantitativeComparisonSmall2003}  stellar dust curves are typically assumed. The Galactic extinction curve \citep{cardelliRelationshipInfraredOptical1989} is commonly assumed for the nebular attenuation curve, consistent in shape in the optical to that derived empirically at $z\sim2$ via multiple Balmer emission lines \citep{reddyMOSDEFSurveyFirst2020}.
    The ratio of two Balmer recombination lines to one another, known as the Balmer decrement, is the standard tool for assessing dust obscuration in ionized gas, and is commonly applied to \ha{} line luminosities (with the Galactic curve) to estimate the intrinsic ionizing photon production \citep[e.g.,][]{shivaeiMOSDEFSurveyDirect2018}.

    Alongside pushing measurements of \ha{} line luminosity and \luv{} to higher redshift, \textit{JWST} has also revealed new information on nebular dust curves in the high-redshift Universe. The ``Assembly of Ultradeep Observations Revealing Astrophysics" Cycle 1 \textit{JWST}/NIRSpec program (AURORA, PID:1914), designed to detect faint auroral emission lines at $z>1.4$, revealed detections of numerous H~\textsc{i} recombination lines in both the Balmer and Paschen series for many of the 97 galaxies targeted by ultradeep spectroscopy. The nebular dust attenuation curve of a galaxy at $z=4.41$ was revealed in \citet{sandersAURORASurveyNebular2025}, finding that it varied significantly from the commonly-assumed Galactic curve, as well as those assumed for stellar attenuation like \citet{calzettiDustContentOpacity2000} and SMC. This analysis was expanded with a sample of 24 AURORA galaxies in \citet{reddyJWSTAURORASurvey2025}, demonstrating the large variability in nebular attenuation curves for galaxies at $z\sim2-4$, and the inconsistency of the average nebular attenuation curve of these objects from other commonly-assumed curves. By leveraging these novel nebular attenuation curves, accurate dust-corrected \ha{} line luminosities can be determined for individual objects, reducing systematics on \xiion{} measurements. In this work, we make use of these curves, as well as utilizing the new AURORA average curve, to explore trends between \xiion{} and galaxy property with minimal nebular dust attenuation systematics.

    The paper is organized as follows. In Section \ref{sec:data}, we review the data products and reduction techniques for the AURORA survey. Section \ref{sec:methods} outlines our procedures for emission-line fitting, spectral energy distribution (SED) modeling, dust attenuation, and estimating \xiion{}, alongside determinations of metallicity. In Section \ref{sec:res}, we report \xiion{} of the AURORA sample and explore how it varies with a number of galaxy properties. We highlight how differences in dust prescriptions affect inferred intrinsic \ha{} line luminosity in \ref{sec:resdust}. We discuss how systematic uncertainties in \ha{} dust correction affect literature results in Section \ref{sec:discussion}, alongside the remaining uncertainties in determining the amount of UV continuum attenuation. We conclude in Section \ref{sec:summary}.

    Throughout this paper, we adopt a standard $\Lambda$CDM cosmology with $\Omega_m$ = 0.3, $\Omega_{\Lambda}$ = 0.7 and $H_0$ = 70 $\textrm{km\,s}^{-1}\textrm{Mpc}^{-1}$. We also employ the AB magnitude system \citep{okeSecondaryStandardStars1983}.
	 
	\section{The AURORA survey} \label{sec:data}
	
	The Assembly of Ultradeep Rest-optical Observations Revealing Astrophysics (AURORA) survey is a Cycle 1 \textit{JWST} program (PID: 1914, Co-PIs: A. Shapley and R. Sanders) that obtained NIRSpec Micro Shutter Assembly (MSA) observations of 97 galaxies, targeting primarily sources at $z=1.4-4.4$ that were estimated to have bright auroral line detections (see \citealt{shapleyAURORASurveyNew2025} for details of the full selection strategy). These sources were in the COSMOS (46) and GOODS-N (51) legacy fields, providing ancillary \textit{HST} and/or \textit{JWST}/NIRCam imaging for each source. 
    We observed each target with the F100LP/G140M, F170LP/G235M, and F290LP/G395M gratings. To ensure uniform line flux detection sensitivity from $1-5\mu$m, we used total integration times of 12.3, 8.0, and 4.2 hrs, respectively. This strategy yielded a typical $3\sigma$ line detection limit of $\rm 5\times10^{-19}\: erg\: s^{-1}\: cm^{-2}$. A three-point nodding pattern was used for these spectroscopic observations. 
	
	\subsection{Reduction and Calibration}
	
	The raw data were processed using the standard STScI NIRSpec reduction pipeline together with custom software developed for the AURORA survey, yielding flat-fielded and wavelength-calibrated two-dimensional (2D) spectrograms. From these, one-dimensional (1D) science and error spectra were optimally extracted using a spatial profile defined by the brightest emission lines in each grating. If no emission lines were present, the integrated continuum profile was used, required for only $\sim1\%$ of grating+target combinations. A full description of survey design and reduction procedures is given in \citet{shapleyAURORASurveyNew2025}.
	
	Given that the NIRSpec MSA microshutter is small relative to the typical size of targeted galaxies, a significant fraction of the light falls outside of the aperture, leading to ``slit loss". This slit loss was corrected using a combination of the wavelength-dependent point-spread function (PSF) and the intrinsic size and morphology of the galaxy as estimated from space-based imaging, with the full details of this procedure described in \citet{reddyPaschenlineConstraintsDust2023a} and \citet{reddyJWSTAURORASurvey2025}.
	
	Final flux calibration ensured that line fluxes and continuum flux densities in overlapping regions were consistent. As described in detail in \citet{sandersAURORASurveyNebular2025}, first, for relative flux calibration, G140M and G395M spectra were scaled to the G235M grating spectra. The second step, absolute flux calibration, was achieved by aligning spectral flux densities with existing multi-wavelength photometric measurements.
	
	\subsection{Ancillary data} \label{sec:phot}
	
	Publicly-available photometric data from \textit{JWST}/NIRCam and \textit{HST}/ACS and WFC3 were obtained through the Dawn JWST Archive \citep[DJA;][]{valentinoAtlasColorselectedQuiescent2023,heintzJWSTPRIMALArchivalSurvey2025}. \textit{JWST} imaging was drawn from the PRIMER, JADES, FRESCO, and JEMS programs \citep{donnanJWSTPRIMERNew2024a,eisensteinOverviewJWSTAdvanced2023,oeschJWSTFRESCOSurvey2023,williamsJEMSDeepMediumband2023}, while \textit{HST} imaging was from CANDELS \citep{groginCANDELSCosmicAssembly2011,koekemoerCANDELSCosmicAssembly2011} and 3D-HST \citep{skelton3DHSTWFC3selectedPhotometric2014}.
    In COSMOS, the \textit{HST} coverage includes the F435W, F606W, F814W, F850LP (ACS) and F105W, F125W, F140W, and F160W (WFC3) bands. The NIRCam coverage includes F090W, F115W, F150W, F200W, F277W, F356W, F410M, and F444W. In GOODS-N, the \textit{HST} data include the F435W, F606W, F775W, F814W, F850LP, F105W, F125W, F140W, and F160W bands, while \textit{JWST}/NIRCam include the following bands: F090W, F115W, F150W, F182M, F200W, F210M, F277W, F335M, F356W, F410M, and F444W. For four COSMOS targets lacking \textit{JWST}/NIRCam coverage, photometry from the 3D-HST catalog was adopted. Photometric measurement procedures for the DJA data are described in \citet{valentinoAtlasColorselectedQuiescent2023}.
	
	\subsection{Nebular dust attenuation curves}\label{sec:nebcurves}
	
	Thanks to the ultradeep spectroscopy and wide wavelength coverage of the AURORA survey, multiple Hydrogen recombination lines were observed, including both Balmer and Paschen series lines. At the median redshift of AURORA $z_{\rm med}=2.70$, the wavelength coverage of AURORA from 1 to $5\: \mu$m allowed for, in the best case scenarios, 5$\sigma$ detections of H$\alpha$ to H12, with simultaneous coverage of Pa$\beta$ with detections through Pa16. 
	
	In \citet{sandersAURORASurveyNebular2025}, the authors used 11 H~\textsc{i} line detections to determine the shape of the nebular attenuation of GOODSN-17940, a star-forming galaxy at $z=4.41$ with an extremely young stellar population. \citet[][hereafter, R25]{reddyJWSTAURORASurvey2025} extended this investigation to include 24 $z=1.5-4.4$ galaxies in AURORA with $>5\sigma$ detections of at least five Hydrogen recombination lines, of which at least two must be Paschen lines, and derived individual nebular attenuation curves for each of these objects. In addition, \citetalias{reddyJWSTAURORASurvey2025} derived an average attenuation curve that describes the mean shape and normalization of the nebular dust attenuation of the 24 galaxy sample. The nebular attenuation curve of GOODSN-17940, the individual curves of the larger 24 galaxy AURORA sample, and the average AURORA curve all display a wide range of shapes and normalizations. The average AURORA curve is inconsistent with that commonly-assumed for nebular dust attenuation, the Galactic extinction curve  \citep{cardelliRelationshipInfraredOptical1989}, as well as other curves used primarily for corrections for starlight such as SMC \citep{gordonQuantitativeComparisonSmall2003} and \citet{calzettiDustContentOpacity2000}. The average AURORA curve has a notably steeper near-IR wavelength dependence and a higher normalization ($R_{\rm V}=6.96$) than the Galactic ($R_{\rm V}=3.1$), SMC ($R_{\rm V}=2.74$), and \citet{calzettiDustContentOpacity2000} ($R_{\rm V}=4.05$) curves. We note that the AURORA curves extend only to the blue optical range, limited by the wavelength coverage of the Balmer lines, and are only extrapolated to the ultraviolet for GOODSN-17940 due to inferences based on its very young age.

	\section{Methodology} \label{sec:methods}
	
	We accurately measure intrinsic \ha{} line fluxes using nebular attenuation curves for individual objects as derived in \citetalias{reddyJWSTAURORASurvey2025} to inform measurements of \xiion. In addition to intrinsic \ha{} luminosity measurements, an estimate of \xiion{} also requires an estimate of \luv{} with a separate stellar dust correction, as determined by SED fitting to broadband photometry.
	
	\subsection{Emission line fluxes and SED Fitting} \label{sec:lineflux}
	
	Emission-line fluxes were measured by fitting Gaussian profiles. Single Gaussians were used for isolated lines and multiple Gaussians for lines in close wavelength proximity. For the initial fits, the continuum level was defined using the best-fit SED model to the multi-wavelength photometry (see description below), prior to applying any corrections of emission-line and/or nebular continuum contributions. These preliminary line fluxes were then used to correct the photometry, after which a new best-fit SED model to the corrected photometry provided the continuum for refitting all lines and deriving the final fluxes. As the continuum is tied to the best-fit SED model, the final measurements of line fluxes such as \ha{} incorporate the effects of underlying stellar absorption. Full details of the line fitting process can be found in \citet{sandersAURORASurveyNebular2025,sandersAURORASurveyHighRedshift2025}.
	
	As described by \citet{sandersAURORASurveyNebular2025} and \cite{shapleyAURORASurveyNew2025}, stellar population properties and continuum models were derived by fitting the corrected photometry with FAST \citep{kriekUltraDeepNearInfraredSpectrum2009} using FSPS models \citep{conroyPropagationUncertaintiesStellar2009a} and a \citet{chabrierGalacticStellarSubstellar2003} IMF, fixing each galaxy's redshift to that measured by AURORA spectroscopy. Delayed-$\tau$ star formation histories ($\textrm{SFR}\propto t\:\textrm{exp}(-t/\tau)$) were assumed, where $t$ is the time since star formation began and $\tau$ the characteristic timescale. Following the approach of \citet{reddyHDUVSurveyRevised2018}, two stellar metallicity-attenuation combinations were tested: a super-solar model ($\zstar=1.4\zsun$) with the \citet{calzettiDustContentOpacity2000} curve, and a sub-solar model ($\zstar=0.27\zsun$) with the SMC extinction curve of \citet{gordonQuantitativeComparisonSmall2003}. For each galaxy, the model yielding the lower $\chi^2$ was adopted, resulting in 76 galaxies fit with the Calzetti prescription and 18 with the SMC curve. 
	These fits provided estimates of star formation rates (SFRs), ages, stellar continuum reddening ($E(B-V)_{\rm stellar}$), and \mstar{}.
	
	Fitting stellar continuum emission also requires a correction for contributions from nebular continuum emission, alongside nebular line emission. This continuum emission was modeled using Cloudy photoionization models \citep{ferland2017ReleaseCloudy2017} tied to the measured \hb{} flux. See \citet{sandersAURORASurveyNebular2025} for the full details of the nebular continuum correction.
	
	\subsection{Ionizing photon production efficiency sample} \label{sec:selection}
	
	As \xiion{} requires both an estimate of intrinsic \ha{} line strength and intrinsic \luv{}, we include AURORA star-forming galaxies with \ha{} detected and photometric constraints near rest-frame 1500\AA{}, while also requiring multiple Balmer lines detections and a well-constrained SED fit in order to determine \ebmvneb{} and $E(B-V)_{\rm stellar}$, respectively. Of the 97 targets first presented in \citet{shapleyAURORASurveyNew2025}, we removed two objects with no recovered redshift, two objects lacking robust photometric SED sampling, five objects with evidence of AGN activity based on broad Balmer-line emission and/or \nii{}$\:/\:$\ha{}$\:> 0.5$ \citep{shapleyJWSTNIRSpecBalmerline2023a,shapleyAURORASurveyNew2025}, and three additional galaxies identified as quiescent. One object, COSMOS-4622, did not have photometry available below rest-frame 2000\AA{}, thus lacked a constraint on \luv{}, and was subsequently removed from the sample. COSMOS-4622 had an nebular dust curve derived in \citetalias{reddyJWSTAURORASurvey2025}.	
	We additionally required significant (3$\sigma$) detections of the Balmer lines of \ha{} and \hb{} in order to determine $E(B-V)_{\rm neb}$, removing 16 objects. Finally, we examined the line ratios of \ha{}/\hb{}, \hb{}/\hg{}, \hg{}/\hd{}, and \hd{}/\he{} where significant at the 3$\sigma$ level, and identified five objects with evidence of non-case B recombination ratios and/or lack of robust flux calibration, and removed them from our sample.
	These cuts result in a final analysis sample of 63 objects. Of these 63 objects, 23 have individual nebular dust curves as derived in \citetalias{reddyJWSTAURORASurvey2025}, which we define as the \aindv{} sample.
	
	\subsection{$L_{\textrm{H}\alpha\textrm{,int}}$ and the nebular dust correction} \label{sec:int_ha}

    Accurate intrinsic \ha{} line luminosities, or $L_{\textrm{H}\alpha\textrm{,int}}$, require robust attenuation corrections, which in turn require multiple Balmer and Paschen lines detections. For the 23 objects in the \aindv{} sample, we use the individual nebular attenuation curves from \citetalias{reddyJWSTAURORASurvey2025} to correct the observed \ha{} luminosities. We use the full suite of Hydrogen recombination lines detected at the $\geq3\sigma$ level to derive \ebmvneb{}.
	
	As in R25, we define the observed flux $f$ for an emission line with a centroid at wavelength $\lambda$ as:
	
	\begin{equation}
		f(\lambda) = f_0(\lambda)\times 10^{-0.4 \ebmvneb k_{\rm neb}(\lambda)},
	\end{equation}
	where $f_0(\lambda)$ is the intrinsic flux of the emission line, \ebmvneb{} is the nebular reddening, and $k_{\rm neb}(\lambda)$ is the value of the assumed nebular attenuation curve (in this case, an individual curve from \citetalias{reddyJWSTAURORASurvey2025}). We define
	
	\begin{equation} \label{eqn:rdef}
	R \equiv \textrm{log}_{10}[\dfrac{f(\lambda_1)}{f(\lambda_2)}] - \textrm{log}_{10}[\dfrac{f_0(\lambda_1)}{f_0(\lambda_2)} ],
	\end{equation}
	such that
	\begin{equation} \label{eqn:rfit}
		R = -0.4 E(B-V)_{\rm neb} [k_{\rm neb}(\lambda_1) - k_{\rm neb}(\lambda_2)].
	\end{equation}
	We calculate R for all significantly detected Balmer and Paschen line fluxes using Equation \ref{eqn:rdef}, setting \ha{} as the normalization in the numerator ($\lambda_1=\lambda_{\textrm{\ha}}=6564.61\textrm{\AA}$). We use intrinsic flux ratios from PyNeb \citep{luridianaPyNebNewTool2015} assuming Case B recombination, with electron density $n_e=100\:\textrm{cm}^{-3}$ and temperature $T=15,000\:\textrm{K}$. Electron densities and temperatures have been directly constrained for a subset of AURORA objects using the [S\textsc{ii}] doublet \citep{toppingAURORASurveyEvolution2025} and auroral emission lines \citep{sandersAURORASurveyHighRedshift2025}, respectively, but we use a single, fixed value for consistency with the derivation of the nebular attenuation curves in \citetalias{reddyJWSTAURORASurvey2025}.
	The assumed electron number density is consistent within a factor of $\sim3$ of electron densities of the bulk of the AURORA objects, and the assumed temperature is consistent within a factor of $\sim1.2$ of their median ionic temperatures. Within PyNeb, intrinsic line ratios show negligible dependence on electron density over the range spanned by AURORA, and vary by less than 5\% for a 20\% change in temperature \citepalias{reddyJWSTAURORASurvey2025}.

\begin{figure*}%
		\centering
			\includegraphics[width=\textwidth]{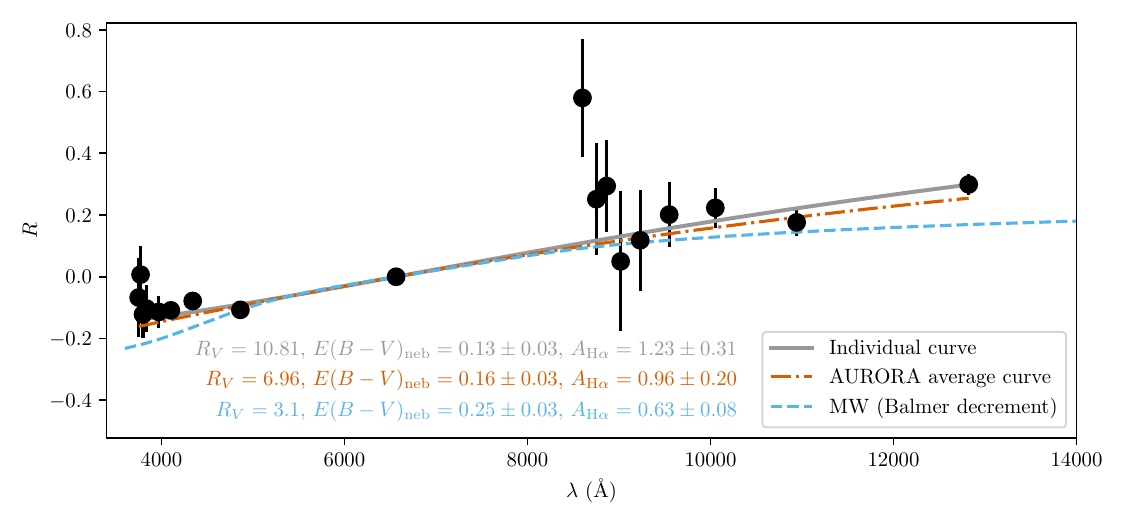}
			\caption{
				Estimation of reddening for GOODSN-22384 ($z=2.993$).
				The logarithm of observed line flux ratios $R$ relative to \ha{}, normalized by the intrinsic line ratio, are shown as a function of the wavelength of a significantly detected line.
                Solid, black points represent $R$ for a suite of Balmer and Paschen recombination lines for this object. These $R$ values are fit by assuming the nebular attenuation curve derived for this object in R25, shown as the grey curve. The red curve shows an alternate fit assuming the average AURORA nebular attenuation curve. In blue, we display the \citet{cardelliRelationshipInfraredOptical1989} curve with \ebmvneb{} determined using only the \ha{}/\hb{} ratio. Total attenuation in magnitudes on \ha{} is reported as $A_{\ha}$, which differs depending on the method for constraining reddening. We also list attenuation curve normalizations $R_V$ and fitted \ebmvneb{} values for each method.
				}
			\label{fig:ebmvfit}
		\end{figure*}

    For each object in \aindv{}, \ebmvneb{} is determined by fitting the set of $R$, with uncertainties propagated from line fluxes, with Equation \ref{eqn:rfit} with \verb|scipy.optimize.curvefit|.
	For a small number of objects, the \ha{}/\hb{} ratio implies a negative \ebmvneb{}, that is, the observed \ha/\hb$\:<2.788$. For these few cases, we calculate the weighted average of \ebmvneb{} as estimated from the Balmer decrement and find it to be 0.03 mag below zero, indicating that there is minimal persistent systematic uncertainty in constraining \ebmvneb{}. We therefore add 0.03 to the uncertainty on \ebmvneb{} determined by the fitting process in quadrature for every object in our analysis sample. This small systematic uncertainty may be driven by imprecise relative flux calibration, slit losses, line fitting techniques, and/or divergence from our Case B recombination assumption.
	We show an \ebmvneb{} fit to $R(\lambda)$ for a single object in Figure \ref{fig:ebmvfit}. As this object, GOODSN-22384, is within the \aindv{} sample, the fiducial fit utilizes the curve derived specifically for GOODSN-22384 in \citetalias{reddyJWSTAURORASurvey2025}, which is shown in grey. The total attenuation of \ha{} for this object is $\aha = \ebmvneb k_{\rm neb}=1.23\pm0.31$ mag.
	
	For objects within the full analysis sample that lack individual attenuation curves ($N=40$), we use the average AURORA nebular attenuation curve to perform the same fitting process using Equations \ref{eqn:rdef} and \ref{eqn:rfit}. We similarly add 0.03 in quadrature to the error on $\ebmvneb$ determined by \verb|scipy.optimize.curvefit|. While not the fiducial fit for this object, we display the fit to the AURORA average curve for GOODSN-22384 in red in Figure \ref{fig:ebmvfit}. For completeness, we also show the \citet{cardelliRelationshipInfraredOptical1989} curve with \ebmvneb{} determined by the \ha{}/\hb{} ratio in light blue, highlighting the typical method of determining total nebular attenuation on \ha{} line luminosities in previous studies. For this object, the variation in \aha{} among the three methods underscores the sensitivity of the inferred intrinsic \ha{} luminosities to dust assumptions. Both the adopted curve shape (and corresponding \ebmvneb{}) and its normalization ($R_V$) influence the resulting \aha{} values. These differences are important to understand as \aha{} is required to estimate the intrinsic \ha{} luminosity, and thus \xiion{} for our sample.

	\subsection{\luv{} and the stellar dust correction} \label{sec:luv}
	
	In contrast to nebular attenuation, stellar attenuation applies to the stellar continuum contributed to by all stars, as opposed to the light emitted only by ionized regions. Thus, stellar and nebular attenuation curves may differ depending on how dust is distributed relative to stars of different ages, as well as any differences between dust grain size and/or column density along lines of sight to different regions of the stellar population. Attenuation of nebular lines has been demonstrated to be significantly higher than that of the stellar continuum in local galaxies \citep{calzettiDustExtinctionStellar1994,wildEmpiricalDeterminationShape2011}. At $z\sim1-1.5$, a similar relationship has been seen, with the caveat that galaxies with large sSFR (and thus very young stellar populations) have similar nebular and stellar attenuation \citep{priceDirectMeasurementsDust2014,puglisiDustAttenuation12016}, although an opposite trend has been found within the MOSFIRE Deep Evolution Field survey at $z\sim2$ \citep{reddyMOSDEFSURVEYMEASUREMENTS2015}.
	
	We estimate the intrinsic non-ionizing UV luminosity following the method of \citet{pahlSpectroscopicAnalysisIonizing2025}. Briefly, the rest-frame UV magnitude at $\lambda_{\rm rest}=1500$\AA{} was derived from the best-fit SED models by integrating the flux density over a 100\AA{} window and converting it to an absolute AB magnitude. Uncertainties were estimated by adopting the average photometric errors near 1500\AA{}, ensuring the error on \luv{} would be empirically characterized by the photometric data. As mentioned in Section \ref{sec:selection}, objects were included in the analysis only if photometric data extended below 2000\AA{} and they had robustly sampled SED. To account for stellar dust effects, the measurements were corrected using the best-fit extinction parameter from the SED fits, applying either a Calzetti or SMC attenuation law depending on which resulted in a lower $\chi^2$ (see Section \ref{sec:lineflux}). We adopt these two curves rather than the AURORA curves from \citetalias{reddyJWSTAURORASurvey2025} for two reasons. First, the Calzetti and SMC curves are known to describe the attenuation of stellar light in $z\sim2$ galaxies, and can reproduce IRX-$\beta$ relations consistent with predictions \citep{reddyHDUVSurveyRevised2018}. Second, the AURORA curves are limited by the wavelength coverage of their Balmer emission and cannot be reliability extrapolated to the UV without assuming an intrinsically blue UV slope, expected only for very young, low-metallicity stellar populations \citep[e.g.][]{reddyMOSDEFSURVEYMEASUREMENTS2015,toppingSearchingExtremelyBlue2022,cullenUltravioletContinuumSlopes2024}. We discuss the use of an individual AURORA UV dust curve for one galaxy with evidence of a very young stellar in Section \ref{sec:res}, and discuss the impact of different stellar dust curves on sample-averaged \xiiono{} values in Section \ref{sec:opticthick}.
	
	\subsection{\xiion{} and uncertainty estimation}
	
	The efficiency of Lyman continuum photon production, \xiion{}, is defined as the intrinsic ionizing photon output ($\dot{n}_{\rm ion}$) per unit non-ionizing UV luminosity density:
	\begin{equation}
		\xiion = \dfrac{\dot{n}_{\rm ion}}{L_{\rm UV,int}} [\textrm{erg}\:\textrm{Hz}^{-1}].
	\end{equation}
	Assuming Case B recombination, a gas temperature of $10,000$K, and an electron density of $100\:\textrm{cm}^{-3}$, $\dot{n}_{\rm ion}$ can be related to the intrinsic $L_{\textrm{H}\alpha\textrm{,int}}$ following the prescription of \citet{leithererSyntheticPropertiesStarburst1995}:
	\begin{equation}
		\dot{\rm n}_{\rm ion} [\textrm{s}^{-1}] = \dfrac{1}{1.36} \times 10^{12} L_{\textrm{H}\alpha\textrm{,int}} [\textrm{erg}\:\textrm{s}^{-1}].
	\end{equation}
	This relation is insensitive to stellar population assumptions. 
    Although the gas temperature adopted here differs from that assumed in deriving the nebular attenuation ($T_e = 15,000\:\textrm{K}$), we use $T_e = 10,000\:\textrm{K}$ to enable direct comparison with literature results, as this value is typically assumed for converting $L_{\textrm{H}\alpha,\textrm{int}}$ to $\dot{n}_{\rm ion}$ \citep[e.g.,][]{shivaeiMOSDEFSurveyDirect2018,pahlSpectroscopicAnalysisIonizing2025,llerenaIonizingPhotonProduction2025}. The intrinsic ionizing output is largely insensitive to electron temperature \citep{leithererSyntheticPropertiesStarburst1995}.
    Throughout this work, we present \xiiono{}, which assumes no escape of ionizing photons; a finite escape fraction would raise the derived value by a factor of 1/(1-\fesc{}).
	
	To quantify measurement errors, we performed 10,000 Monte Carlo realizations, similar to the method of \citet{pahlSpectroscopicAnalysisIonizing2025}. In each iteration, the observed UV flux density was perturbed within its uncertainty and converted to an intrinsic luminosity using a randomly selected \av{} drawn from the FAST posterior. Simultaneously, the \ha{} luminosity and nebular reddening were varied within their error distributions. Any negative perturbed \ebmv{} values were set to 0. The resulting \xiiono{} distribution was summarized by its median value, with the 16th and 84th percentiles adopted as the confidence interval.
	
	\subsection{Metallicities}
	
	Thanks to the detection of auroral emission lines, including \oiiia{}, \oiia{}, \siiia{}, and \siia{}, electron temperatures and direct-method metallicities were determined for AURORA galaxies \citep[for full details see][]{sandersAURORASurveyHighRedshift2025}. Out of the  63 (23) in the full analysis (\aindv{}) sample, 33 (19) objects have direct metallicities available which we draw from Table 2 of \citet{sandersAURORASurveyHighRedshift2025}. As a proxy for both ionization parameter and, indirectly, metallicity, we also computed the line ratio $\textrm{O32}=$\oiiif{}$/$\oii{}, performing dust corrections via \ebmvneb{} as determined in Section \ref{sec:int_ha}.
	
	\section{Ionizing photon production efficiency in AURORA} \label{sec:res}
	
\begin{figure*}%
		\centering
		\includegraphics[width=\textwidth]{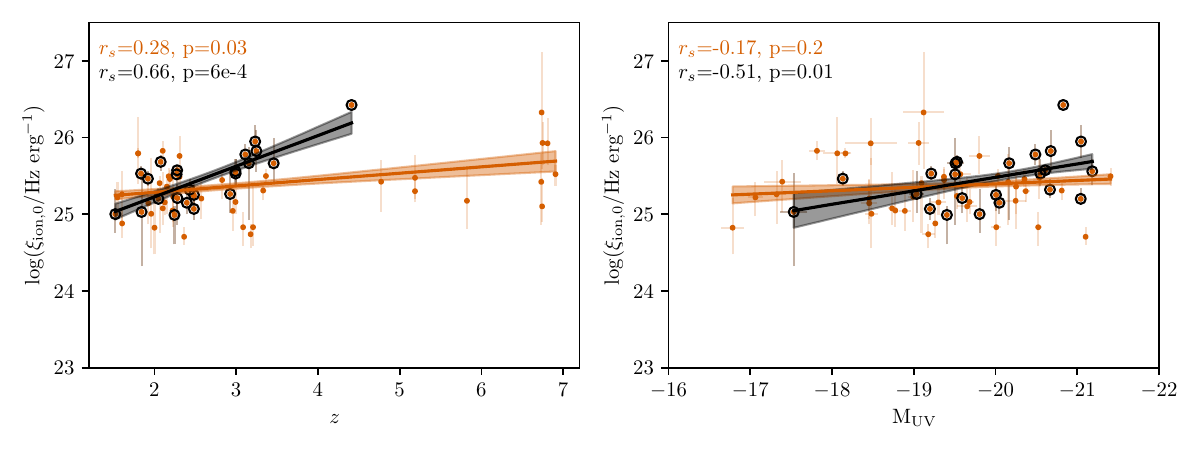}
		\caption{Ionizing photon production efficiency as a function of redshift and \muv{} within AURORA. Datapoints highlighted with black, concentric circles are within the \aindv{} subsample, which each have individual nebular attenuation curves. The rest of the analysis sample uses the AURORA average curve for determining \ebmvneb{}. Spearman correlation coefficients are shown in orange for the full sample and black for the \aindv{} subsample. Linear fits and 1$\sigma$ confidence intervals to the full analysis sample are shown as dark orange curves with light orange shaded regions, while fits to \aindv{} are shown as black curves with grey shaded regions. Spearman correlation tests and linear regression reveals elevated $\xiiono$ at high redshift, and mild, positive evolution with \luv{}.
		}
		\label{fig:xiion_best}
	\end{figure*}

	We present \xiiono{} of AURORA galaxies as a function of redshift and \muv{} in Figure \ref{fig:xiion_best}. We highlight objects in the \aindv{} sample with black, concentric circles, as these objects have individually-determined nebular attenuation curves and thus have the most accurate corrections for nebular dust attenuation. Within \aindv{}, we calculate the Spearman correlation coefficient between \xiiono{} and redshift, finding $r_s=0.66$ with $p=6\times10^{-4}$, demonstrating a highly significant, positive correlation. Datapoints in orange have fewer significantly-detected Hydrogen recombination lines (and/or have less than two detected Paschen lines), and thus do not have individually-determined nebular attenuation curves. We used the average AURORA curve for these objects to determine nebular attenuation.  For the full analysis sample (including both \aindv{} and the objects using the average AURORA curve), we again calculate the Spearman correlation coefficient, finding $r_s=0.28$ with $p=0.03$, a weaker but still significant ($p<0.05$, $>2\sigma$) trend. While the inclusion of galaxies outside of the \aindv{} sample increases the dynamic range in both \xiiono{} and redshift, such systems have less-precise determinations of \xiiono{}. The positive trend we have recovered between \xiiono{} and redshift has been well established by \textit{JWST} analyses using both spectroscopy and photometry \citep{boyettExtremeEmissionLine2024a,simmondsLowmassBurstyGalaxies2024,simmondsIonizingPropertiesGalaxies2024,pahlSpectroscopicAnalysisIonizing2025,begleyEvolutionIIIHv2025}, extending to $z\sim6$ and beyond.
	
	Repeating this analysis for the trend between \xiiono{}	and \muv{}, we find $r_s=-0.51$ and $p=0.01$ within \aindv{}, demonstrating a positive correlation between \xiiono{} and \luv{} at a $2\sigma$ significance level, such that more UV luminous galaxies have larger \xiiono{} values. Within the full analysis sample, the correlation coefficient reduces to $r_s=-0.17$ with $p=0.2$ indicating a lack of significant relationship between the two variables. 
	
	While the Spearman correlation tests monotonicity and is robust to outliers, linear regression encodes uncertainties on each measurement, which are still significant for our \xiiono{} estimates primarily due to an uncertain stellar dust correction (see discussion in Section \ref{sec:uncertainty}). To perform linear fits between \xiion{} and galaxy property, we use the method of \citet{sandersDirectEbasedMetallicities2024}, which is appropriate for fitting trends in which measurement uncertainty is smaller than the intrinsic scatter between the variables. We generate 10,000 perturbed versions of the dataset based on their uncertainties and fit each trend using \verb|scipy.stats.linregress|. From these 10,000 trend lines, we compute median \xiiono{} across a grid of galaxy property, then perform a final fit to these medians to determine the final fitting coefficients. Confidence intervals on the best-fit relation are derived from the 10,000 realizations.
	This method prevents a bias towards a few objects with very small measurement uncertainty, which is important as \xiiono{} uncertainties vary widely across our sample. Fitting the full  analysis sample, we find:
	\begin{equation}
		\textrm{log}(\xi_{\rm ion,0}/\textrm{Hz erg}^{-1})=(0.08\pm0.03)\times z +25.12\pm0.09
	\end{equation}
	and
	\begin{equation} \label{eqn:xiion_muv}
		\textrm{log}(\xi_{\rm ion,0}/\textrm{Hz erg}^{-1})=(-0.05\pm0.03)\times \muv +24.5\pm0.7.
	\end{equation}
	These two relationships hint at positive relationships between both \xiiono{} and $z$ and \xiiono{} and \luv{} at the $2\sigma$ and $1\sigma$ significance level, respectively, when performing linear fits. Fitting the \aindv{} subsample, we find:
    \begin{equation}
        \textrm{log}(\xi_{\rm ion,0}/\textrm{Hz erg}^{-1})=(0.40\pm0.08)\times z+24.43\pm0.21
    \end{equation}
    and
    
    \begin{equation}
        \textrm{log}(\xi_{\rm ion,0}/\textrm{Hz erg}^{-1})=(-0.18\pm0.08)\times \textrm{M}_{\rm UV}+21.97\pm1.65
    \end{equation}
    showing more significant ($4\sigma$ and $2\sigma$, respectively) trends when examining objects only with the most reliable \aha{} determinations. However, these trends remain limited by the dynamic range of the \aindv{} subsample and are more susceptible to outliers. 
    Additionally, galaxies within the \aindv{} subsample show higher mean \xiiono{}, driven by both their higher median Balmer decrement (reflecting selection for non-negligible reddening) and their higher median observed \ha{} luminosity due to the requirement of multiple Paschen-line detections \citepalias{reddyJWSTAURORASurvey2025}.
    
    The \xiiono{}-\muv{}-$z$ relationships found in the full AURORA analysis sample follow that which was found by a large, spectroscopic analysis across early-release \text{JWST} surveys \citep{pahlSpectroscopicAnalysisIonizing2025}, and are consistent with slopes found in the \textit{JWST} Advanced Deep Extragalactic Survey via photometry \citep{simmondsIonizingPropertiesGalaxies2024}. However, studies that did not apply a Balmer-decrement-based nebular dust correction have reported the opposite trend between \xiiono{} and \luv{}, with UV-faint galaxies exhibiting higher \xiiono{}, in contrast to our findings \citep{llerenaIonizingPhotonProduction2025}. We compare our methodology to studies from literature that find negative trends between \xiiono{} and \luv{} in Section \ref{sec:litcompare}.
	
\begin{figure*}
	\centering
	\includegraphics[width=\textwidth]{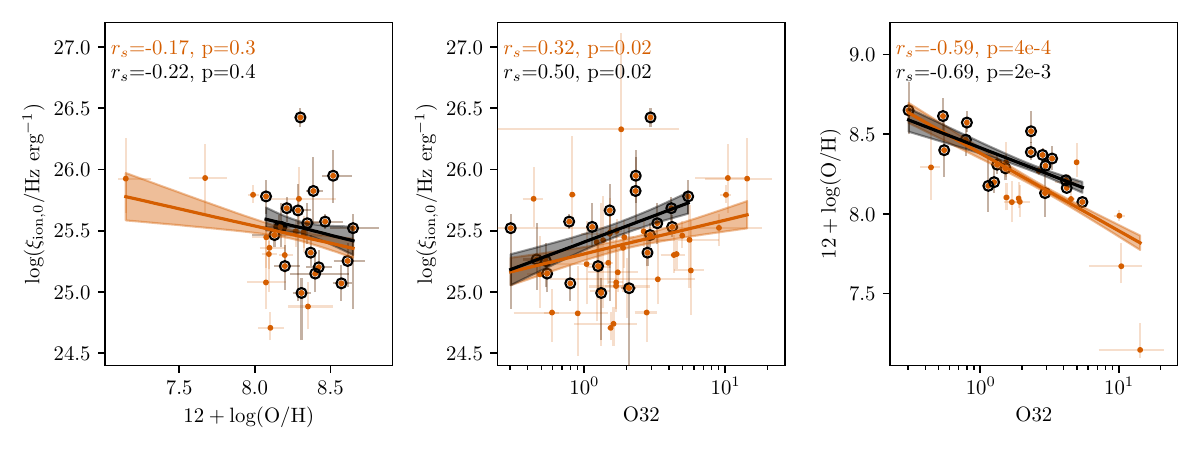}
	\caption{Relationships between ionizing photon production efficiency, O32, and oxygen abundance within AURORA. Datapoints, trends, and Spearman correlations are colored as in Figure \ref{fig:xiion_best}. Oxygen abundances are determined via the direct method, utilizing detections of temperature-sensitive auroral emission lines. The Spearman correlation test reveals no strong trend between \xiiono{} and \oh{}, despite significant correlations between \xiiono{} and O32 and anti-correlations between \oh{} and O32. Linear regression reveals a marginally significant ($1\sigma$), negative trend between \xiiono{} and \oh{}. Expanding the sample to include additional lower-metallicity objects is required to fully probe this trend.
	}
	\label{fig:xiion_Z_O32}
	\end{figure*}

	Thanks to the detection of faint auroral lines within AURORA, the sample presents a unique opportunity to study both \xiiono{} with a precise nebular dust correction and ``direct" method metallicities at high redshift. We display \xiiono{} as a function of \oh{} in the left panel of Figure \ref{fig:xiion_Z_O32}. Despite strong anti-correlations found between these parameters when inferring \oh{} from strong-line calibrations \citep[e.g.,][]{shivaeiMOSDEFSurveyDirect2018,pahlSpectroscopicAnalysisIonizing2025,llerenaIonizingPhotonProduction2025}, we do not recover any significant trend between \xiiono{} and direct-method metallicity when using the Spearman correlation test. Linear regression, which is more sensitive to the two objects in our analysis sample with the lowest metallicity, implies a marginally significant ($1\sigma$), negative slope:
	\begin{equation}
		\begin{split}
		\textrm{log}(\xi_{\rm ion,0}/\textrm{Hz erg}^{-1})=(-0.28\pm0.18)\times[12+\textrm{log}(\textrm{O/H})] \\ +27.77\pm1.45.
		\end{split}
	\end{equation}
    A slope consistent with zero is seen within the \aindv{} sample.
	Combinations of stellar and photoionization models predict that lower-metallicity stellar populations have harder ionizing spectra and higher ionizing photon production efficiencies \citep[e.g.,][]{stanwayStellarPopulationEffects2016,bylerNebularContinuumLine2017}, fundamentally motivating the strong trends seen in literature between \xiion{} and the equivalent width of \oiii{}, as well as with O32 and \oiii{}/\hb{} \citep[e.g.,][]{chevallardPhysicalPropertiesHionizingphoton2018,reddyHDUVSurveyRevised2018,tangMMTMMIRSSpectroscopy2019,shivaeiMOSDEFSurveyDirect2018,pahlSpectroscopicAnalysisIonizing2025,llerenaIonizingPhotonProduction2025}. Indeed, within AURORA we find strong trends between \xiiono{} and O32, via both the Spearman correlation test and linear regression, shown in the middle panel of Figure \ref{fig:xiion_Z_O32}, as well as a strong trend between \oh{} and O32, shown in the right panel. This analysis reveals a large scatter between \xiiono{} and direct-method \oh{}, with tentative trends driven by a few objects at $\oh \lesssim 8.0$. An analysis including additional objects with low metallicities will definitively test the strength of this correlation.
	
\begin{figure*}
		\centering
		\includegraphics[width=\textwidth]{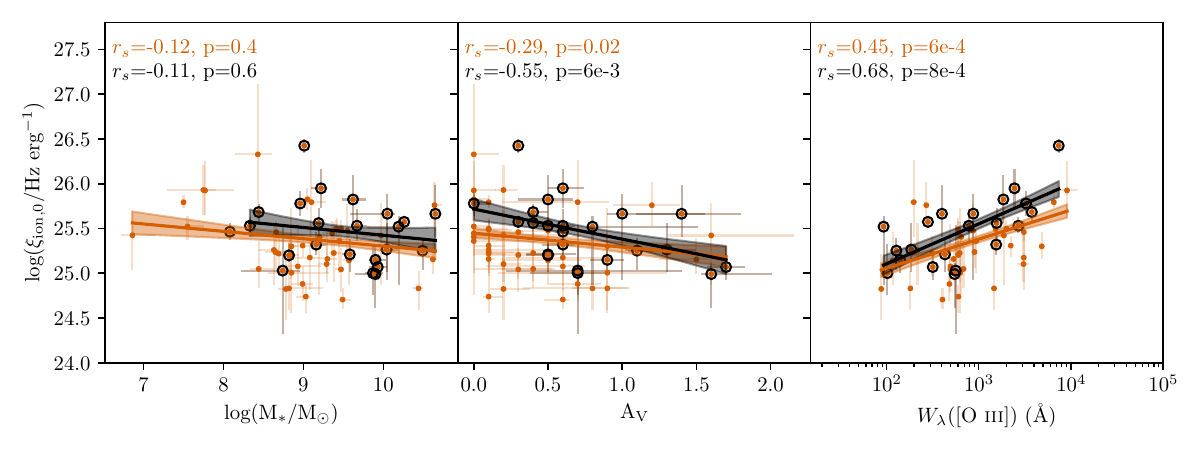}
		\caption{Relationships between \xiiono{} and stellar mass \mstar{}, stellar $A_V$, and \oiiif{} equivalent width within AURORA. Datapoints, trends, and Spearman correlations are colored as in Figure \ref{fig:xiion_best}. We find strong trends between \xiiono{} and \av{} and \xiiono{} and \woiii{}, with the Spearman correlation test showing no correlation between \xiiono{} and \mstar{}. \woiii{} is an effective indirect tracer of \xiiono{}, and the calibration derived here from AURORA galaxies has minimal nebular dust systematics.
 		}
		\label{fig:xiion_props}
	\end{figure*}

	We display relationships between \xiiono{} and additional galaxy properties of interest in Figure \ref{fig:xiion_props}, including \mstar{}, \av{}, and the equivalent width of \oiiif{} (\woiii{}). Spearman correlation tests between \xiiono{} and \mstar{} do not reject the null hypothesis of statistical independence, within both the \aindv{} and full analysis samples. Linear regression reveals a slope consistent with zero within \aindv{}, and a slightly negative correlation within the full sample, with a slope of $-0.08\pm0.05$ between logarithmic quantities. This slope is notably lower than trends found within other \textit{JWST} studies \citep{castellanoIonizingPhotonProduction2023,chenUnveilingCosmicReionization2024,begleyEvolutionIIIHv2025,llerenaIonizingPhotonProduction2025}. We find a statistically-significant negative correlation between \xiiono{} and \av{}, with the strongest trend revealed within the \aindv{} sample with Spearman $r_s=-0.55$ and $p=6\times10^{-3}$. Performing a linear regression between \xiiono{} and \av{} within the full analysis sample, we find a slope of $-0.15\pm0.09$, demonstrating a negative relationship at $1\sigma$ significance, and a negative relationship at $2\sigma$ significance within \aindv{} ($-0.33\pm-0.15$). Finally, we recover the strong trend between \xiiono{} and \woiii{} in both \aindv{} and the full analysis sample that is ubiquitous across samples spanning redshift and galaxy property \citep{chevallardPhysicalPropertiesHionizingphoton2018,reddyHDUVSurveyRevised2018,tangMMTMMIRSSpectroscopy2019,pahlSpectroscopicAnalysisIonizing2025,llerenaIonizingPhotonProduction2025}. This trend is
	\begin{equation}
		\begin{split}
        \textrm{log}(\xi_{\rm ion,0}/\textrm{Hz erg}^{-1})=(0.33\pm0.07)\times \textrm{log}(\woiii{}\:(\mbox{\normalfont\AA}))\\+24.40\pm0.20
		\end{split}
	\end{equation}
	within the full analysis sample. This diagnostic has the benefit of being determined with \xiiono{} estimates with minimal dust systematics, as well as \woiii{} estimated purely spectroscopically from significant detection of the continuum in the NIRSpec grating spectra (as opposed to a combination of spectroscopy and photometry for continuum estimation), and is appropriate for application to other samples within a similar redshift range and span of galaxy properties.

	We note that one galaxy, GOODSN-17940, has an estimated $\lxiiono=26.42_{-0.08}^{+0.08}$, which exceeds the theoretical limit of $\xiion\sim26.0$ predicted by spectral synthesis models of Pop II and Pop III stellar populations \citep{masedaElevatedIonizingPhoton2020,nanayakkaraReconstructingObservedIonizing2020,lecroqNewPrescriptionSpectral2025}. GOODSN-17940 is  a starburst galaxy dominated by a young stellar population $<10\:\textrm{Myr}$ in age and extreme nebular emission line equivalent widths and strengths. 
    \citet{sandersAURORASurveyNebular2025} argue that, for this object, the lines of sight to the nebular and stellar continuum emission may be similar due to its young stellar population, implying comparable reddening. Applying \ebmvneb{} to the observed \luv{} with the attenuation curve derived in \citet{sandersAURORASurveyHighRedshift2025} yields an adjusted value of $\lxiiono=25.37$, consistent with theoretical limits. This object directly illustrates the importance of constraining the UV continuum attenuation curve for star-forming galaxies at high redshift. We note that using this alternate \xiiono{} value for GOODSN-17940 has a minor impact on the trends presented in this section.
	
	\section{Dust prescriptions and \aha{}} \label{sec:resdust}
	
	Critical to determinations of \xiiono{} is the intrinsic line luminosity of \ha{}, which can be converted to an intrinsic ionizing photon output without sensitivity to stellar population assumptions and minimal sensitivity to electron temperature and density \citep[e.g.,][]{leithererSyntheticPropertiesStarburst1995,bylerNebularContinuumLine2017}. This intrinsic line luminosity requires an estimate of nebular line attenuation (\aha{}), which, as we have shown for one object in AURORA in Figure \ref{fig:ebmvfit}, depends strongly on the shape and normalization of the assumed nebular attenuation curve. \aha{} is similarly important for other quantities that depend on intrinsic \ha{} line luminosity, such as SFR. The derivation of individual nebular attenuation curves for AURORA objects allow for an estimation of \aha{} with minimal systematics, allowing for a ``ground truth" \aha{} to which other dust prescriptions can be compared.
	
\begin{figure}
		\centering
		\includegraphics[width=\columnwidth]{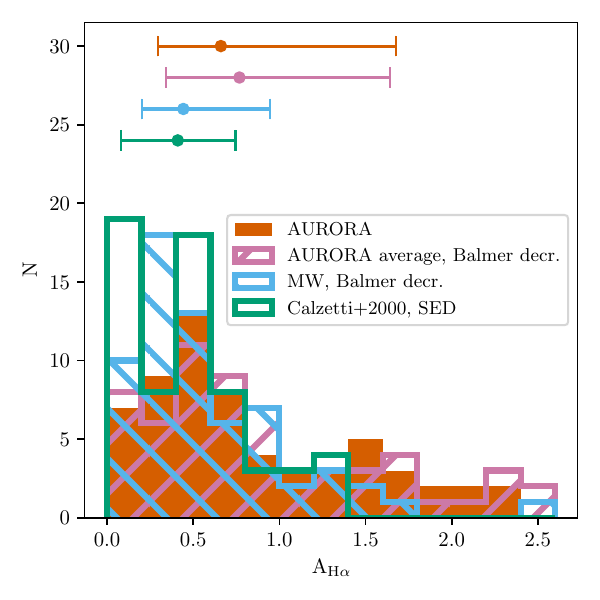}
		\caption{The distribution of the total attenuation in magnitudes on \ha{}, \aha{}, for various dust prescriptions. In orange, the fiducial \aha{} values are displayed, determined using individual dust curves for objects in \aindv{} and the average AURORA curve for all other objects, using simultaneous fits to significantly-detected H~\textsc{i} recombination line ratios. In pink, the AURORA average curve was used for all objects, with \ebmvneb{} determined from the Balmer decrement. In blue, the Milky Way \citep{cardelliRelationshipInfraredOptical1989} extinction curve was used with a Balmer decrement to determine \ebmvneb{}. In green, \aha{} values were determined directly from fits to stellar population synthesis models, with the \citet{calzettiDustContentOpacity2000} stellar attenuation curve assumed. Medians and 16th and 84th percentiles of the distribution are shown as solid points with errorbars. Within AURORA, the average AURORA curve and Balmer decrement result in a similar \aha{} distribution as our fiducial analysis, while using the Milky Way curve or \aha{} derived from the SED result in lower median \aha{} values (resulting in biased \xiiono{}).
		}
		\label{fig:aha_hist}
	\end{figure}

	In Figure \ref{fig:aha_hist}, we present the distribution of \aha{} as determined by the \ebmv{} fits described in \ref{sec:int_ha} for the full AURORA sample (orange histogram), with median and 16th and 84th percentiles of $\aha =0.66_{-0.37}^{+1.02}$. This distribution contains \aha{} values calculated for 23 objects in \aindv{}, utilizing individual dust attenuation curves as derived in \citetalias{reddyJWSTAURORASurvey2025}, and 40 objects for which the average AURORA curve was used and \aha{} was derived by fitting all detected Hydrogen recombination lines. In pink, we display the distribution of \aha{} values that would be derived if using the average AURORA curve for all objects, with \ebmvneb{} determined with the Balmer decrement (\ha{}/\hb{}). The median of this distribution is $\aha=0.77_{-0.43}^{+0.89}$, with the shape and median comparable to our fiducial \aha{} determination. These values represent what one would infer for \aha{} if only detections of \ha{} and \hb{} were available to estimate the Balmer decrement, and the average AURORA nebular attenuation curve was assumed, empirically derived in \citetalias{reddyJWSTAURORASurvey2025}. In contrast, in blue, we show the distribution of \aha{} values derived using the Galactic extinction curve \citep{cardelliRelationshipInfraredOptical1989} and a Balmer decrement based on \ha{}/\hb{}, a common method for estimating the attenuation on nebular lines at high redshift \citep[e.g.,][]{wildEmpiricalDeterminationShape2011,dominguezDustExtinctionBalmer2013,reddyMOSDEFSURVEYMEASUREMENTS2015,shivaeiMOSDEFSurveyDirect2018,shapleyJWSTNIRSpecBalmerline2023a,clarkeStarFormingMainSequence2024}. 
    As the Galactic curve differs in both shape and normalization relative to the AURORA curves, the median $\aha=0.44_{-0.24}^{+0.52}$ based on this third method is lower than our fiducial determination, and would result in lower average \xiiono{} and SFR values derived from $L_{\textrm{H}\alpha\textrm{,int}}$. Finally, we calculate \aha{} using the \citet{calzettiDustContentOpacity2000} stellar attenuation curve and deriving \ebmvst{} from fits to emission-line corrected photometry (see Section \ref{sec:lineflux}). If we assume $\ebmvst=\ebmvneb$, then the median $\aha$ is significantly lower than in our fiducial analysis: $0.41_{-0.33}^{+0.34}$. This method may be employed if no nebular attenuation information can be gleaned from the available observations, such as a lack of spectroscopy or a single Hydrogen recombination line detection.

\begin{figure*}
		\centering
		\includegraphics[width=\textwidth]{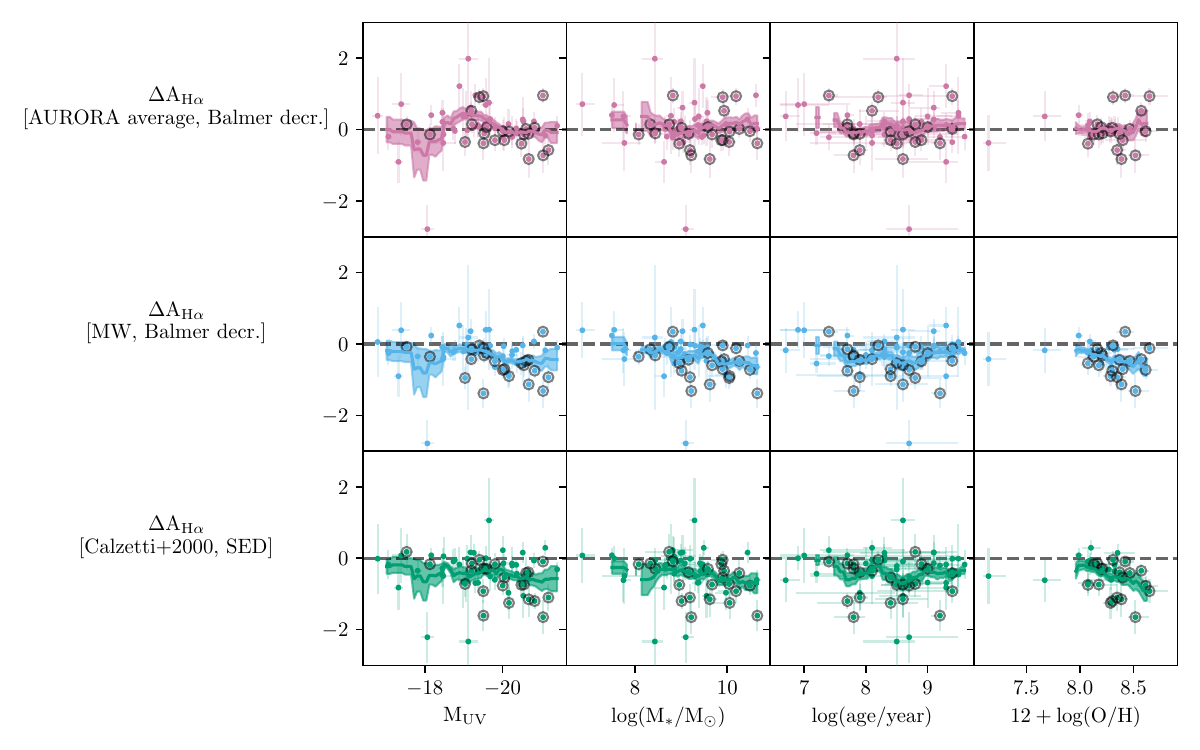}
		\caption{
			The difference between \aha{} as determined by our fiducial analysis and alternate dust prescriptions within AURORA, as function of galaxy property. Each row corresponds to a dust prescription also shown in Figure \ref{fig:aha_hist}. Running means of $\Delta \aha$ are shown as solid curves, with standard error on the mean shown as shaded regions around the curves. A window size of the dynamic range in galaxy property divided by 12 was used to calculate the running mean. Datapoints highlighted with black, concentric circles are within the \aindv{} subsample, which each have individual nebular attenuation curves. Using the average AURORA curve does not introduce systematics compared to our ``ground-truth" attenuation determination, but alternate assumptions result in preferentially-lower intrinsic \ha{} luminosities for UV bright, low-mass, high-metallicity galaxies.
		}
		\label{fig:aha_grid}
	\end{figure*}

    We find that the median \aha{} for a galaxy sample changes depending on the nebular dust prescription used. Moreover, how these prescriptions vary \textit{with galaxy property} could affect measurements from dust-corrected line luminosities in unexpected ways. In Figure \ref{fig:aha_grid}, we present the difference in \aha{} for individual galaxies in AURORA between the different derivation methods described above. Here, our fiducial \ebmvneb{} fits using either individual curves or the average curve from AURORA are the reference value within each comparison. We show the difference in \aha{} when using an alternate \ebmvneb{} derivation as a function of a number of galaxy properties: \muv{}, \mstar{}, stellar age, and \oh{}. When estimating \aha{} from the average AURORA nebular attenuation curve and \ha{}/\hb{}, we find no structure in the residual of $\Delta \aha$. That is, \aha{} with the AURORA curve and \ha/\hb{} is the same on average as the fiducial \aha{} determination as a function of galaxy property. Thus, measurements derived from dust-corrected line luminosities (such as \xiion{}) and subsequent trends with galaxy property will not be biased using this method.
	
	When using \aha{} estimated using the Galactic extinction curve and a Balmer decrement measurement, \aha{} values are on average lower than those from our fidicial analysis, and, \aha{} values are preferentially lower at bright \muv{}, high \mstar{}, and high oxygen abundance. These biases can be seen through apparent structure in the residual between Galactic \aha{} values and those from our fiducial analysis. Thus, using the Galactic curve as opposed to the average AURORA curve or individually-determined nebular attenuation curves will produce $L_{\textrm{H}\alpha\textrm{,int}}$ values that are biased low for UV bright, high stellar mass, high metallicity objects, which will directly propagate to biases in \xiiono{}. Similar biases are seen when comparing \aha{} derived from SED fitting and our fiducial \aha{} estimates. When assuming $\ebmvneb{}=\ebmvst{}$ and a \citet{calzettiDustContentOpacity2000} curve, \aha{} values are preferentially lower at bright \muv{}, high \mstar{}, and high oxygen abundance. We note that these biases are minimized when the observed \ha{}/\hb{} ratio approaches the intrinsic ratio, that is, when there is evidence of little nebular reddening.
	
	\section{Discussion} \label{sec:discussion}
	
	\subsection{Dust assumptions in literature measurements of \xiion{}} \label{sec:litcompare}
	
	Within AURORA, we find evidence for positive relationships between \xiiono{} and redshift, \luv{}, \woiii{}, and O32, and negative relationships between \xiion{} and stellar $A_V$ and metallicity, through a combination of Spearman correlation test and linear regression. Our sample contains 23 objects with nebular dust attenuation curves determined on an individual object basis. As dust curves can vary significantly from object to object, this determination minimizes systematics on \ebmvneb{}, and thus \xiion{}. The sample additionally contains 40 objects without individual nebular dust attenuation curves, for which we implement an average AURORA curve derived from the individual curves.
	
	Literature measurements of \xiion{} using \textit{JWST} implement dust corrections in a variety of ways. In \citet{pahlSpectroscopicAnalysisIonizing2025}, a larger sample of 163 star-forming galaxies at $z\sim2-7$ was examined from the JWST Advanced Deep Extragalactic Survey \citep[JADES;][]{eisensteinOverviewJWSTAdvanced2023} and the Cosmic Evolution Early Release Science Survey \citep[CEERS;][]{finkelsteinCompleteCEERSEarly2023}, requiring detections of \ha{} and \hb{} for Balmer decrement determinations of \ebmvneb{} assuming a Galactic extinction curve. Using identical methods for intrinsic \luv{} measurements as in this work, a significant, positive correlation between \xiion{} and \luv{} was found.  
    
\begin{figure}
		\centering
		\includegraphics[width=\columnwidth]{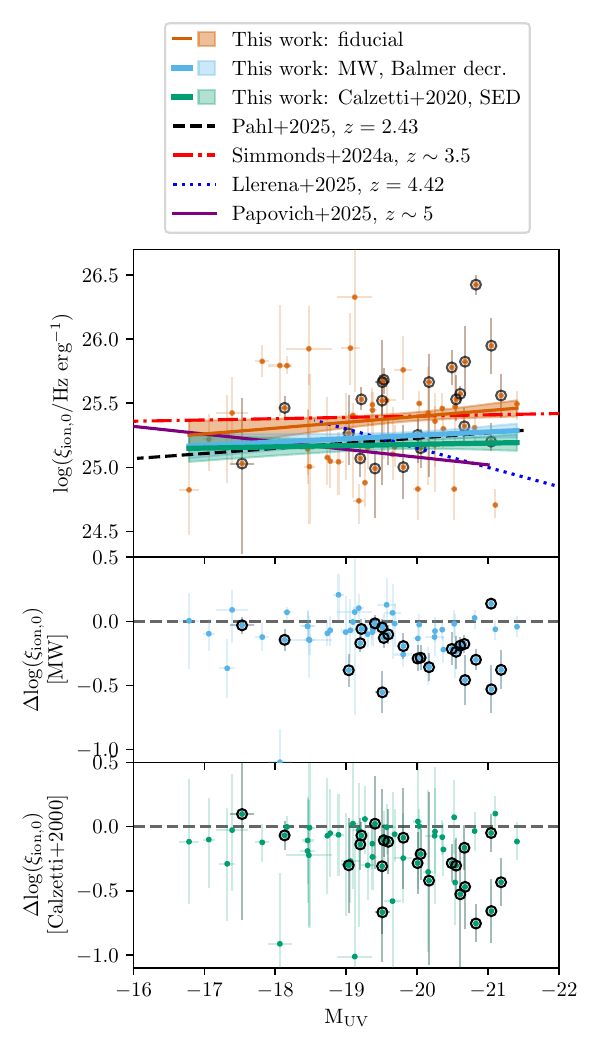}
		\caption{Comparison of AURORA and literature \xiiono{} estimates, folding in differences in dust assumptions. \textbf{Top:} Fiducial measurements of \xiiono{} are shown as data points, colored identically to Figure \ref{fig:xiion_best}. The solid, orange line displays the best-fit trend between \xiiono{} and \muv{}. In the solid, blue curve, the best-fit trend between \xiiono{} and \muv{} is shown for measurements of AURORA objects in which the Galactic \citet{cardelliRelationshipInfraredOptical1989} extinction curve was assumed for nebular attenuation, with \ebmvneb{} measured from the Balmer decrement. In the solid, green curve, the best fit trend between \xiiono{} and \muv{}  is shown for measurements of AURORA objects in which \ebmvneb{} was assumed to be the same as \ebmvst{}.  Additional \xiiono{}-\muv{} curves from literature are also shown, with the caveat that the \citet{papovichGalaxiesEpochReionization2025} trend was determined from \xiion{} with non-zero \fesc{}.
        \textbf{Middle:} The difference between fiducial \xiiono{} measurements and those with the Milky Way assumption are shown as solid, blue points. Galaxies in the \aindv{} are highlighted with black, concentric circles. \textbf{Bottom:} The difference between fiducial \xiiono{} measurements and those with the SED-based \ebmvneb{} assumption are shown as solid, green points in the bottom panel. Galaxies in the \aindv{} are highlighted with black, concentric circles. 
		}
		\label{fig:xiion_dx}
	\end{figure}

    In Figure \ref{fig:xiion_dx}, we again display the trend between \xiion{} and \muv{} within the full AURORA analysis sample. Below, in blue points, we display the difference in \xiion{} we would have measured if we implemented Balmer decrement based determinations of \ebmvneb{} with the Galactic extinction curve, similar to the methods of \citet{pahlSpectroscopicAnalysisIonizing2025}. We see, on average, \xiion{} values are lower when implementing the Milky Way curve, expected from the average differences in \aha{} between these two methods presented in Section \ref{sec:resdust}. We also see that the difference in \xiiono{} between the our fiducial measurements and the Galactic-based measurements are larger at brighter \muv{}. This reduces the slope of the trend between \xiiono{} and \muv{} to be consistent with zero within AURORA: the slope of the best-fit linear regression is $-0.03\pm0.03$. The \xiiono{}-\luv{} relation from \citet{pahlSpectroscopicAnalysisIonizing2025} is consistent with the AURORA Milky Way \xiiono{} within the uncertainties (shown as the blue, shaded region in Figure \ref{fig:xiion_dx}), demonstrating that the AURORA sample recovers a similar \xiiono{}-\muv{} relation as the JADES and CEERS combined sample when assuming the same dust curve.
	
	In \citet{llerenaIonizingPhotonProduction2025}, a larger-still sample of 761 star-forming galaxies at $z\sim4-10$ was compiled from JADES, CEERS, and GLASS \citep{treuGLASSJWSTEarlyRelease2022} surveys, requiring a single, spectroscopic detection of \ha{} and/or \hb{} in order to increase sample size, and to push measurements beyond $z\gtrsim7$ where \ha{} shifts out of the range of \textit{JWST}/NIRSpec. Thus, nebular dust corrections were performed using stellar population synthesis fits to broadband photometry, deriving \ebmvst{} using the \citet{calzettiDustContentOpacity2000} stellar attenuation law and applying them to Balmer line fluxes to determine \xiiono{}. We similarly compute \xiiono{} assuming $\ebmvneb=\ebmvst$ from our SED fits detailed in Section \ref{sec:lineflux}, for which \citet{calzettiDustContentOpacity2000} was assumed, and show the difference in \xiiono{} to our fiducial results in green points in Figure \ref{fig:xiion_dx}. On average, \xiiono{} values are lower using SED-based \aha{}, and show a more negative deviation at brighter \muv{}. The resulting linear trend fit to AURORA \xiiono{} with SED-based \aha{} has a slope consistent with zero. The trend found within the analysis of \citet{llerenaIonizingPhotonProduction2025} is inconsistent with our trend, demonstrating that while the assumption of SED-based \ebmvneb{} does induce a more negative trend between \xiiono{} and \luv{}, it does not fully describe the strongly negative relationship found within the JADES+CEERS+GLASS spectroscopic sample presented in \citet{llerenaIonizingPhotonProduction2025}.
	
	Additional factors must be examined to fully explain the discrepancy in trends reported between \xiiono{} and \luv{} across different works. In \citet{simmondsLowmassBurstyGalaxies2024}, a mass-complete sample of galaxies drawn from JADES was analyzed, with \xiiono{} inferred from SED fits to broadband photometry via the fitting code \textsc{prospector} \citep{johnsonStellarPopulationInference2021}. During the fitting process, a two-component dust model was assumed, with a variable dust index applying to young and old stellar populations \citep{kriekDustAttenuationLaw2013}. The flat trend between \xiiono{} and \muv{} found by \citet{simmondsLowmassBurstyGalaxies2024} within a similar redshift subsample as AURORA is shown in Figure \ref{fig:xiion_dx}, and is comparable to our fiducial trend. This consistency is present despite differences in average dust curve normalization between AURORA curves and that of \citet{kriekDustAttenuationLaw2013}, where AURORA curves tend to have higher average $R_V$ values. 
	
    In contrast, \citet{papovichGalaxiesEpochReionization2025} examined galaxies from both JADES and CEERS at $z\sim4.5-9$, implementing spectrophotometric fits to broadband photometry and \textit{JWST}/NIRSpec prism data to constrain \xiion{} and \fesc{} simultaneously with the fitting code \textsc{bagpipes} \citep{carnallInferringStarFormation2018}. The nebular dust attenuation was assumed to be the same as the stellar dust attenuation, assuming a \citet{calzettiDustContentOpacity2000} dust curve. When comparing AURORA \xiiono{} with a similar dust prescription to the \xiion{}-\muv{} found in \citet{papovichGalaxiesEpochReionization2025}, we find similar average values but inconsistent slopes: \xiiono{} within AURORA when assuming a stellar \ebmv{} derived with the \citet{calzettiDustContentOpacity2000} curve has no significant dependence on \luv{}, while \citet{papovichGalaxiesEpochReionization2025} finds a strong, negative trend with \luv{}. Recent work analyzing the \textit{JWST}/NIRSpec Early eXtragalactic Continuum and Emission Line Science \citep[EXCELS;][]{carnallJWSTEXCELSSurvey2024} survey with the assumption of a Galactic nebular dust curve reports a weak slope between \xiiono{} and \muv{} of $0.08\pm0.04$, mildly inconsistent with AURORA under the same nebular dust assumption \citep{begleyJWSTEXCELSSurvey2025}. These works altogether indicate that while differences in the treatment of nebular dust attenuation can explain some differences in $\xiion{}-\muv{}$ relationships, sample selection, data reduction, or stellar population synthesis modeling methods must be examined to explain our lack of agreement. In future work, we will focus on this last point by examining \xiion{} in AURORA as estimated by various photometric and spectrophotometric codes.
		
	\subsection{Optically-thick obscuration of ionizing photons and applicability to reionization models} \label{sec:opticthick}
	A primary physical interpretation of the variation among AURORA nebular dust attenuation curves, which differ in shape and normalization from standard prescriptions when constrained by Balmer and Paschen line detections, is that they point to non-unity dust covering fractions in star-forming galaxies at cosmic noon and beyond \citepalias{reddyJWSTAURORASurvey2025}. This covering fraction can be invoked to explain, for example, the inability for the Galactic extinction curve to fit both Balmer and Paschen line ratios for GOODSN-22384 as shown in Figure \ref{fig:ebmvfit} with the assumption of a foreground dust screen. 
    With a non-unity dust covering fraction, a portion of the Ha emission comes from relatively unreddened sightlines, while the remaining originates from reddened sightlines that could be optically-thick to Balmer emission.
    Optically-thick \ha{} emission is only revealed when using curves derived from longer-wavelength Paschen lines, which are relatively less affected by dust attenuation. This obscuration is the primary origin of the larger average $R_V$ values of AURORA nebular dust attenuation curves, and thus, the larger \aha{} derived, and larger \xiiono{} values when using the AURORA curves. 
    
    For our fiducial analysis, the median \xiiono{} within AURORA is $\lxiiono{}=25.32_{-0.28}^{+0.44}$, with $\lxiiono=25.47_{-0.22}^{+0.61}$ at $z\geq4$. These fiducial \xiiono{} measurements are higher than recent works at similar redshifts that use the Galactic extinction curve \citep{shivaeiMOSDEFSurveyDirect2018,pahlSpectroscopicAnalysisIonizing2025,begleyEvolutionIIIHv2025} or \citet{calzettiDustContentOpacity2000} curve \citep{llerenaIonizingPhotonProduction2025,papovichGalaxiesEpochReionization2025}, and have a stronger, positive relationship between \xiiono{} and redshift than is seen in a photometric study of a mass complete sample of JADES galaxies \citep{simmondsIonizingPropertiesGalaxies2024}. Our values are comparable to those recovered by samples selected as extreme emission line galaxies \citep[e.g.,][]{boyettExtremeEmissionLine2024a,simmondsLowmassBurstyGalaxies2024}, which initially motivated the ``ionizing photon budget crisis" presented in \citet{munozReionizationJWSTPhoton2024a}. Nonetheless, when assuming a $\fesc$ prescription directly motivated by the $z\sim3$ Keck Lyman Continuum Spectroscopic survey \citep[$\fesc=0.09$ at $\muv>-21$, $\fesc=0$ otherwise;][]{steidelKeckLymanContinuum2018,pahlUncontaminatedMeasurementEscaping2021}, reionization modeling following the methods of \citet{pahlSpectroscopicAnalysisIonizing2025} shows that assuming a relatively high, constant $\lxiiono = 25.47$ from AURORA still results in reionization completing at $z\sim 6$.
	
	As \xiiono{} is an \textit{efficiency} of ionizing photon production, revealing a population of obscured ionizing photon production requires an investigation of whether the UV emission has a similar obscured portion. This obscuration is not currently constrained by SED fitting of rest-UV to rest-NIR photometry. 
    To assess this, we compare SFRs derived from $L_{\mathrm{H}\alpha}$, corrected for optically thick obscuration, with SFRs dervied from \luv{}, where the UV luminosity is corrected for dust attenuation using $A_{\rm UV}$ from SED fitting. Under an SMC stellar dust curve and subsolar stellar metallicity stellar population, SFR(UV) is on average $\sim0.2$ dex lower than SFR(\ha{}). In contrast, assuming a \citet{calzettiDustContentOpacity2000} stellar dust curve and near-solar metallicity yields SFR(UV) values that are $\sim0.2$ dex higher than SFR(\ha{}) (Reddy et al., in prep). These offsets illustrate the scale of adjustment that may be needed to bring our individually dust-corrected \ha{} luminosities (the numerator of \xiiono{}) into consistency with estimates of the intrinsic \luv{} (the denominator).

    While comparisons between SFR(\ha{})–SFR(UV) are useful, SED fitting does not explicitly account for an optically thick component to the UV continuum, so this type of relative scaling may not be fully reliable. Assuming an SMC stellar dust curve for \luv{} and our fiducial nebular dust correction, the AURORA sample yields a median $\lxiiono = 25.57_{-0.28}^{+0.48}$, increasing to $\lxiiono = 25.78_{-0.27}^{+0.53}$ for objects at $z > 4$. If we reduce \luv{} by 0.2 dex to align SFR(\ha{}) and SFR(UV), such that $\lxiiono$ is assumed to be a fixed value of 25.58 at $z>4$, reionization modeling in an identical manner to that of \citet{pahlSpectroscopicAnalysisIonizing2025} predicts reionization completing at $z\sim6.2$. Alternatively, adopting a Calzetti stellar dust curve for all AURORA galaxies gives a median $\lxiiono = 25.31_{-0.28}^{+0.46}$ (and $\lxiiono = 25.47_{-0.23}^{+0.52}$ for $z > 4$). Decreasing \luv{} by 0.2 dex for consistency between the two SFR estimates (and thus decreasing $\lxiiono$ to 25.67) causes the same reionization modeling to end at $z \sim 6.6$, later than implied by independent constraints on the neutral fraction. Quantifying these systematic uncertainties related to stellar dust attenuation is therefore essential for reliable use of \xiiono{} in reionization models.

    A complete understanding of stellar dust systematics within AURORA will require direct constraints on the optically thick obscuration of UV emission. Constraints on dust emission via IR photometry allow for an estimate of the total intrinsic UV luminosity based on energy balance arguments. Within energy-balance SED fitting, the integrated infrared luminosity from dust heated by stars is consistent with the amount of stellar light absorbed by dust from the UV to the near-IR. Even with a single reliable flux point, SED+L$_{\rm IR}$ fitting can be reliably implemented to constrain stellar dust attenuation \citep{salimDustAttenuationCurves2018}. Observations of the dust continuum emission with facilities such as ALMA and NOEMA could constrain L$_{\rm IR}$, providing crucial constraints on the degree of stellar UV continuum attenuation. In addition, the 7.7 $\mu$m PAH feature contains roughly half of the total PAH luminosity \citep{smithMidInfraredSpectrumStarforming2007} and is tightly correlated with $L_{\rm IR}$ at $z\sim2$ \citep{ronayneCEERS77Mm2024,shivaeiTightCorrelationPAH2024}. \textit{JWST}/MIRI photometry of AURORA targets could efficiently constrain the strength of the 7.7 $\mu$m PAH feature, and thus $L_{\rm IR}$, while simultaneously exploring the correlation between nebular dust attenuation curve shape/normalization and PAH emission strength. Longer-wavelength observations would thus significantly improve systematic uncertainties on the UV continuum dust correction.
	
	\subsection{Remaining systematic uncertainty in determining \xiion{}} \label{sec:uncertainty}
	
	The depth and wavelength coverage of the \textit{JWST}/NIRSpec observations of AURORA objects result in high precision line flux measurements and significant detections of H~\textsc{i} recombination lines in the Paschen series, allowing for individual nebular attenuation curves to be constrained and reduction in systematic uncertainty on the nebular dust attenuation correction to $L_{\mathrm{H}\alpha}$. To quantify this reduction, we compute the mean absolute difference in \xiiono{} when implementing alternate dust prescriptions as compared to our fiducial dust analysis, $\langle |\Delta \lxiiono| \rangle$. Under the assumption of the AURORA average curve for all objects and the derivation of \ebmvneb{} via the Balmer decrement, we find $\langle |\Delta \lxiiono| \rangle = 0.148$, representing the average amount of systematic uncertainty when assuming a single nebular attenuation curve and a single H~\textsc{i} line ratio, even if that curve describes the sample on average. When assuming a curve that does not describe the sample, such as the Galactic extinction curve for AURORA galaxies, this systematic uncertainty is only slightly higher ($\langle |\Delta \lxiiono| \rangle = 0.165$), but the \xiiono{} values and trends with galaxy property such as \muv{} are systematically biased (see Figure \ref{fig:xiion_dx}).
	
	Despite this reduction in systematic uncertainty, the typical estimate of \xiiono{} within AURORA still has appreciable statistical and measurement uncertainty. The mean log-relative error of \xiiono{} within our analysis sample is $\langle \delta^{\mathrm{eff}} \rangle = 0.244$, comparable to the difference in \xiiono{} at the the extreme ends of our \muv{} distribution using the best-fit linear regression reported in Equation \ref{eqn:xiion_muv}. Considering the importance of higher-fidelity stellar dust corrections discussed in the previous section, we wish to quantify which sources contribute to our stated errorbars on \xiiono{}.	
	To this end, we calculate the mean log-relative error of different quantities that contribute to our stated measurement uncertainties. These quantities include measurement errors on \ha{} luminosity, error on \aha{} due to line-ratio uncertainty and the fitting of \ebmvneb{}, photometric uncertainty near rest-frame 1500\AA{} (informing error on \luv{}), and the uncertainty of $\textrm{A}_{\textrm{UV}}$ from the SED fitting process. For quantities with asymmetric errorbars, each lower and upper error was first averaged before the mean was computed for the full sample:
	\begin{equation}
		\langle \delta^{\mathrm{eff}} \rangle = \frac{1}{N}\sum_{i=1}^N \delta_i^{\mathrm{eff}} = \frac{1}{2N}\sum_{i=1}^N \bigl(\delta_{i,+} + \delta_{i,-}\bigr),
	\end{equation}
	where $\delta_{i,+}$ and  $\delta_{i,-}$ are the upper and lower errors in log space:
	\begin{equation}
		\delta_{i,+} = \log\!\left(\frac{x_i + \sigma_{i,+}}{x_i}\right),\qquad
		\delta_{i,-} = \log\!\left(\frac{x_i}{x_i - \sigma_{i,-}}\right).
	\end{equation}
	for a measurement $x_i$ with asymmetric error $\sigma_{i,+}$ and $\sigma_{i,-}$. These mean log-relative errors, presented as a dimensionless number, are tabulated in Table \ref{tab:err}. 
			\begin{table} 
		\centering
		\caption{Mean log-relative errors for quantities required for the estimation of \xiiono{}}
		\begin{tabular}{lr}
			\toprule
			& $\langle \delta_{\text{eff}} \rangle$ \\
			\midrule
			$\xi_{\rm ion,0}$ & 0.244 \\
			$L_{\mathrm{H}\alpha}$ & 0.007 \\
			$A_{\mathrm{H}\alpha}$ & 0.138 \\
			$L_{\mathrm{UV}}$ & 0.027 \\
			$A_{\mathrm{UV}}$ & 0.191 \\
			\bottomrule
		\end{tabular}
		\label{tab:err}
	\end{table}
	As expected, measurement uncertainties of \ha{} line luminosities are low, given the remarkable depth of the AURORA spectra (average S/N of \ha{} is 143). Similarly, as \luv{} is constrained by \textit{HST} and \textit{JWST} photometry, the measurement error is small compared to the error on \xiiono{}. While appreciable uncertainty does remain in \aha{} driven by fitting \ebmvneb{} with all available H~\textsc{i} recombination line ratios (with higher order lines being relatively faint, even at the depth of AURORA), the largest contributor to our errorbars is the uncertain constraints on $A_{\textrm UV}$ via SED fitting. Inclusion of mid-IR or far-IR photometric measurements can reduce uncertainty on $A_{\textrm UV}$ by an order of magnitude \citep{pacificiArtMeasuringPhysical2023}. A combination of precise nebular dust attenuation and mid-IR/far-IR constraints on dust continuum and PAH emission will enable the most precise \xiiono{} measurements to date from AURORA galaxies.
		
		\section{Summary} \label{sec:summary}
		
		In this work, we analyze star-forming galaxies within the AURORA survey in order to constrain their ionizing-photon production efficiencies (\xiion{}). Due to the depth and wavelength coverage of AURORA, precise nebular attenuation curves for 24 objects have been determined, as well as an average curve that describes the full sample (R25). We analyze a sample of 63 objects within AURORA, including those with and without individual nebular attenuation curves, to make estimates of \xiiono{} that have minimal nebular dust systematics. We investigate the effects of different dust prescriptions on median \xiiono{} values within AURORA, as well as trends with galaxy property such as \muv{}, \mstar{}, stellar age, and \oh{}. Our main results are summarized below.
		
		\begin{enumerate}
			\item We find significant, positive trends between \xiiono{} and redshift, O32, and \woiii{}, and significant, negative trends between \xiiono{} and A$_{\rm V}$ within AURORA via the Spearman correlation test. Within a sample that contains only objects with individually-determined nebular attenuation curves, we find a positive trend between \xiiono{} and \luv{} at the $2\sigma$ level, indicating that fainter galaxies are less efficient at producing ionizing radiation. Linear regression reveals a shallow, positive slope between \xiiono{} and \luv{} within the full sample, as well as a marginally-negative trend between \xiiono{} and \mstar{}.
			\item Direct method metallicities have been constrained for AURORA objects thanks to observations of temperature-sensitive auroral lines \citep{sandersAURORASurveyHighRedshift2025}. We present the first exploration of \xiiono{} and direct-method metallicity at these redshifts, find no strong trend between the two parameters via the Spearman correlation test. This lack of correlation is present despite predictions from stellar population models \citep{stanwayStellarPopulationEffects2016,bylerNebularContinuumLine2017}, and strong trends between \xiiono{} and O32 within AURORA. Using linear regression, we find a negative relationship between \xiiono{} and \oh{} at the $1\sigma$ level. We conclude that the lack of dynamic range in metallicity for objects in AURORA may be contributing to the weak evidence of correlation between these parameters.
			\item We examine the effects of nebular dust attenuation on the \ha{} line luminosity, \aha{}, predicted by several dust prescriptions for AURORA galaxies. We find that \aha{} depends strongly on the shape and normalization of the assumed nebular attenuation curve, which varies significantly from galaxy to galaxy  \citepalias{reddyJWSTAURORASurvey2025}. We find that using the average AURORA curve and \ebmvneb{} determined from the Balmer decrement produces similar \aha{} values on average as those from our fiducial analysis, which includes individually-determined nebular attenuation curves, with no bias across galaxy properties. In contrast, using the Galactic extinction curve \citep{cardelliRelationshipInfraredOptical1989} produces lower \aha{} values than our fiducial analysis, and induces a relative bias in \aha{} across \luv{}, \mstar{}, and \oh{}. Similar biases are seen when assuming $\ebmvneb=\ebmvst$.
			\item We find that using alternate dust prescriptions induce a more negative trend between \xiiono{} and \luv{} within AURORA. While choice of dust prescription appear to explain some of the differences in breadth of \xiiono{} vs. \luv{} relationships recovered in the literature \citep[e.g.,][]{simmondsIonizingPropertiesGalaxies2024,pahlSpectroscopicAnalysisIonizing2025,llerenaIonizingPhotonProduction2025,papovichGalaxiesEpochReionization2025}, the range in \xiiono{} values found within AURORA with differing dust assumptions do not fully explain all the observed deviations. We conclude that sample selection or SED modelling differences must be invoked to fully explain these discrepancies.
		\end{enumerate}
		
		While our analysis within has reduced the systematic uncertainty on \xiiono{} via precise nebular dust corrections, significant uncertainty still remains on the stellar dust correction. Simultaneous constraints on the SED from the rest-UV to the rest-IR can constrain the optically-thick obscuration of stellar light, and thus produce \xiiono{} measurements that minimize both types of dust systematics within AURORA.
		\linebreak
		\linebreak
		AJP was generously supported by a Carnegie Fellowship through the Carnegie Observatories while conducting this work. K.G. acknowledges support from Australian Research Council Laureate Fellowship FL180100060.
		This work is based on observations made with the NASA/ESA/CSA James Webb Space Telescope. The data were obtained from the Mikulski Archive for Space Telescopes at the Space Telescope Science Institute, which is operated by the Association of Universities for Research in Astronomy, Inc., under NASA contract NAS5-03127 for JWST. Data were also obtained from the DAWN JWST Archive maintained by the Cosmic Dawn Center.
        The spectroscopic observations used in this paper can be found in MAST: \dataset[10.17909/hvne7139]{doi:10.17909/hvne7139}.

\begin{thebibliography}{}
\expandafter\ifx\csname natexlab\endcsname\relax\def\natexlab#1{#1}\fi
\providecommand{\url}[1]{\href{#1}{#1}}
\providecommand{\dodoi}[1]{doi:~\href{http://doi.org/#1}{\nolinkurl{#1}}}
\providecommand{\doeprint}[1]{\href{http://ascl.net/#1}{\nolinkurl{http://ascl.net/#1}}}
\providecommand{\doarXiv}[1]{\href{https://arxiv.org/abs/#1}{\nolinkurl{https://arxiv.org/abs/#1}}}

\bibitem[{Becker {et~al.}(2021)Becker, D'Aloisio, Christenson, Zhu, Worseck, \&
  Bolton}]{beckerMeanFreePath2021}
Becker, G.~D., D'Aloisio, A., Christenson, H.~M., {et~al.} 2021, Monthly
  Notices of the Royal Astronomical Society, 508, 1853,
  \dodoi{10.1093/mnras/stab2696}

\bibitem[{Begley {et~al.}(2025{\natexlab{a}})Begley, McLure, Cullen, McLeod,
  Dunlop, Carnall, Stanton, Shapley, Cochrane, Donnan, Ellis, Fontana, Grogin,
  \& Koekemoer}]{begleyEvolutionIIIHv2025}
Begley, R., McLure, R.~J., Cullen, F., {et~al.} 2025{\natexlab{a}}, Monthly
  Notices of the Royal Astronomical Society, 537, 3245,
  \dodoi{10.1093/mnras/staf211}

\bibitem[{Begley {et~al.}(2025{\natexlab{b}})Begley, McLure, Cullen, Carnall,
  Stanton, Scholte, McLeod, Dunlop, {Arellano-C{\'o}rdova}, Bondestam, Donnan,
  Hamadouch, Shapley, \& Stevenson}]{begleyJWSTEXCELSSurvey2025}
---. 2025{\natexlab{b}}, The {{JWST EXCELS Survey}}: {{A}} Spectroscopic
  Investigation of the Ionizing Properties of Star-Forming Galaxies at 1,
  arXiv

\bibitem[{Boyett {et~al.}(2024)Boyett, Bunker, {Curtis-Lake}, Chevallard,
  Cameron, Jones, Saxena, Charlot, Curti, Wallace, Arribas, Carniani, Willott,
  Alberts, Eisenstein, Hainline, Hausen, Johnson, Rieke, Robertson, Stark,
  Tacchella, Williams, Chen, Egami, Endsley, Kumari, Laseter, Looser, Maseda,
  Scholtz, Shivaei, Simmonds, Smit, {\"U}bler, \&
  Witstok}]{boyettExtremeEmissionLine2024a}
Boyett, K., Bunker, A.~J., {Curtis-Lake}, E., {et~al.} 2024, Monthly Notices of
  the Royal Astronomical Society, 535, 1796, \dodoi{10.1093/mnras/stae2430}

\bibitem[{Byler {et~al.}(2017)Byler, Dalcanton, Conroy, \&
  Johnson}]{bylerNebularContinuumLine2017}
Byler, N., Dalcanton, J.~J., Conroy, C., \& Johnson, B.~D. 2017, The
  Astrophysical Journal, 840, 44, \dodoi{10.3847/1538-4357/aa6c66}

\bibitem[{Calzetti {et~al.}(2000)Calzetti, Armus, Bohlin, Kinney, Koornneef, \&
  {Storchi-Bergmann}}]{calzettiDustContentOpacity2000}
Calzetti, D., Armus, L., Bohlin, R.~C., {et~al.} 2000, The Astrophysical
  Journal, 533, 682, \dodoi{10.1086/308692}

\bibitem[{Calzetti {et~al.}(1994)Calzetti, Kinney, \&
  {Storchi-Bergmann}}]{calzettiDustExtinctionStellar1994}
Calzetti, D., Kinney, A.~L., \& {Storchi-Bergmann}, T. 1994, The Astrophysical
  Journal, 429, 582, \dodoi{10.1086/174346}

\bibitem[{Cardelli {et~al.}(1989)Cardelli, Clayton, \&
  Mathis}]{cardelliRelationshipInfraredOptical1989}
Cardelli, J.~A., Clayton, G.~C., \& Mathis, J.~S. 1989, The Astrophysical
  Journal, 345, 245, \dodoi{10.1086/167900}

\bibitem[{Carnall {et~al.}(2018)Carnall, McLure, Dunlop, \&
  Dav{\'e}}]{carnallInferringStarFormation2018}
Carnall, A.~C., McLure, R.~J., Dunlop, J.~S., \& Dav{\'e}, R. 2018, Monthly
  Notices of the Royal Astronomical Society, 480, 4379,
  \dodoi{10.1093/mnras/sty2169}

\bibitem[{Carnall {et~al.}(2024)Carnall, Cullen, McLure, McLeod, Begley,
  Donnan, Dunlop, Shapley, Rowlands, Almaini, {Arellano-C{\'o}rdova}, Barrufet,
  Cimatti, Ellis, Grogin, Hamadouche, Illingworth, Koekemoer, Leung, Lovell,
  {P{\'e}rez-Gonz{\'a}lez}, Santini, Stanton, \&
  Wild}]{carnallJWSTEXCELSSurvey2024}
Carnall, A.~C., Cullen, F., McLure, R.~J., {et~al.} 2024, Monthly Notices of
  the Royal Astronomical Society, 534, 325, \dodoi{10.1093/mnras/stae2092}

\bibitem[{Castellano {et~al.}(2023)Castellano, Belfiori, Pentericci,
  Calabr{\`o}, Mascia, Napolitano, Caro, Charlot, Chevallard, Lake, Talia,
  Bongiorno, Fontana, Fynbo, Garilli, Guaita, McLure, Merlin, Mignoli, Moresco,
  Pompei, Pozzetti, Lopez, Saxena, Santini, Schaerer, Schreiber, Shapley,
  Vanzella, \& Zamorani}]{castellanoIonizingPhotonProduction2023}
Castellano, M., Belfiori, D., Pentericci, L., {et~al.} 2023, A\&A, 675, A121,
  \dodoi{10.1051/0004-6361/202346069}

\bibitem[{Chabrier(2003)}]{chabrierGalacticStellarSubstellar2003}
Chabrier, G. 2003, Publications of the Astronomical Society of the Pacific,
  115, 763, \dodoi{10.1086/376392}

\bibitem[{Chen {et~al.}(2024)Chen, Motohara, Spitler, Nakajima, \&
  Terao}]{chenUnveilingCosmicReionization2024}
Chen, N., Motohara, K., Spitler, L., Nakajima, K., \& Terao, Y. 2024, Towards
  Unveiling the {{Cosmic Reionization}}: The Ionizing Photon Production
  Efficiency (\${\textbackslash}xi\_\{ion\}\$) of {{Low-mass
  H}}\${\textbackslash}alpha\$ Emitters at \$z {\textbackslash}sim 2.3\$,
  arXiv.
\newblock \doarXiv{2404.10998}

\bibitem[{Chevallard {et~al.}(2018)Chevallard, Charlot, Senchyna, Stark,
  {Vidal-Garc{\'i}a}, Feltre, Gutkin, Jones, Mainali, \&
  Wofford}]{chevallardPhysicalPropertiesHionizingphoton2018}
Chevallard, J., Charlot, S., Senchyna, P., {et~al.} 2018, Monthly Notices of
  the Royal Astronomical Society, 479, 3264, \dodoi{10.1093/mnras/sty1461}

\bibitem[{Clarke {et~al.}(2024)Clarke, Shapley, Sanders, Topping, Brammer,
  Bento, Reddy, \& Kehoe}]{clarkeStarFormingMainSequence2024}
Clarke, L., Shapley, A.~E., Sanders, R.~L., {et~al.} 2024, The {{Star-Forming
  Main Sequence}} in {{JADES}} and {{CEERS}} at \$z{$>$}1.4\$:
  {{Investigating}} the {{Burstiness}} of {{Star Formation}},  arXiv.
\newblock \doarXiv{2406.05178}

\bibitem[{Conroy {et~al.}(2009)Conroy, Gunn, \&
  White}]{conroyPropagationUncertaintiesStellar2009a}
Conroy, C., Gunn, J.~E., \& White, M. 2009, The Astrophysical Journal, 699,
  486, \dodoi{10.1088/0004-637X/699/1/486}

\bibitem[{Cullen {et~al.}(2024)Cullen, McLeod, McLure, Dunlop, Donnan, Carnall,
  Keating, Magee, {Arellano-Cordova}, Bowler, Begley, Flury, Hamadouche, \&
  Stanton}]{cullenUltravioletContinuumSlopes2024}
Cullen, F., McLeod, D.~J., McLure, R.~J., {et~al.} 2024, Monthly Notices of the
  Royal Astronomical Society, 531, 997, \dodoi{10.1093/mnras/stae1211}

\bibitem[{Dom{\'i}nguez {et~al.}(2013)Dom{\'i}nguez, Siana, Henry, Scarlata,
  Bedregal, Malkan, Atek, Ross, Colbert, Teplitz, Rafelski, McCarthy, Bunker,
  Hathi, Dressler, Martin, \& Masters}]{dominguezDustExtinctionBalmer2013}
Dom{\'i}nguez, A., Siana, B., Henry, A.~L., {et~al.} 2013, Astrophysical
  Journal, 763, \dodoi{10.1088/0004-637X/763/2/145}

\bibitem[{Donnan {et~al.}(2024)Donnan, McLure, Dunlop, McLeod, Magee,
  {Arellano-C{\'o}rdova}, Barrufet, Begley, Bowler, Carnall, Cullen, Ellis,
  Fontana, Illingworth, Grogin, Hamadouche, Koekemoer, Liu, Mason, Santini, \&
  Stanton}]{donnanJWSTPRIMERNew2024a}
Donnan, C.~T., McLure, R.~J., Dunlop, J.~S., {et~al.} 2024, Monthly Notices of
  the Royal Astronomical Society, 533, 3222, \dodoi{10.1093/mnras/stae2037}

\bibitem[{Eisenstein {et~al.}(2023)Eisenstein, Willott, Alberts, Arribas,
  Bonaventura, Bunker, Cameron, Carniani, Charlot, {Curtis-Lake}, D'Eugenio,
  Endsley, Ferruit, Giardino, Hainline, Hausen, Jakobsen, Johnson, Maiolino,
  Rieke, Rieke, Rix, Robertson, Stark, Tacchella, Williams, Willmer, Baker,
  Baum, Bhatawdekar, Boyett, Chen, Chevallard, Circosta, Curti, Danhaive,
  DeCoursey, {de Graaff}, Dressler, Egami, Helton, Hviding, Ji, Jones, Kumari,
  L{\"u}tzgendorf, Laseter, Looser, Lyu, Maseda, Nelson, Parlanti, Perna,
  Pusk{\'a}s, Rawle, Rodr{\'i}guez Del~Pino, Sandles, Saxena, Scholtz, Sharpe,
  Shivaei, Silcock, Simmonds, Skarbinski, Smit, Stone, Suess, Sun, Tang,
  Topping, {\"U}bler, Villanueva, Wallace, Whitler, Witstok, \&
  Woodrum}]{eisensteinOverviewJWSTAdvanced2023}
Eisenstein, D.~J., Willott, C., Alberts, S., {et~al.} 2023, Overview of the
  {{JWST Advanced Deep Extragalactic Survey}} ({{JADES}}),
  \dodoi{10.48550/arXiv.2306.02465}

\bibitem[{Ferland {et~al.}(2017)Ferland, Chatzikos, Guzm{\'a}n, Lykins, {van
  Hoof}, Williams, Abel, Badnell, Keenan, Porter, \&
  Stancil}]{ferland2017ReleaseCloudy2017}
Ferland, G.~J., Chatzikos, M., Guzm{\'a}n, F., {et~al.} 2017, The 2017
  {{Release Cloudy}},  arXiv, \dodoi{10.48550/arXiv.1705.10877}

\bibitem[{Finkelstein {et~al.}(2023)Finkelstein, Leung, Bagley, Dickinson,
  Ferguson, Papovich, Akins, Haro, Dave, Dekel, Kartaltepe, Kocevski,
  Koekemoer, Pirzkal, Somerville, Yung, Amorin, Backhaus, Behroozi, Bisigello,
  Bromm, Casey, Ortiz, Cheng, Chworowsky, Cleri, Cooper, Davis, {de la Vega},
  Elbaz, Franco, Fontana, Fujimoto, Giavalisco, Grogin, Holwerda,
  {Huertas-Company}, Hirschmann, Iyer, Jogee, Jung, Larson, Lucas, Mobasher,
  Morales, Morley, Mukherjee, {Perez-Gonzalez}, Ravindranath, Rodighiero,
  Rowland, Tacchella, Taylor, Trump, \&
  Wilkins}]{finkelsteinCompleteCEERSEarly2023}
Finkelstein, S.~L., Leung, G. C.~K., Bagley, M.~B., {et~al.} 2023, The
  {{Complete CEERS Early Universe Galaxy Sample}}: {{A Surprisingly Slow
  Evolution}} of the {{Space Density}} of {{Bright Galaxies}} at z
  {\textasciitilde} 8.5-14.5,  arXiv, \dodoi{10.48550/arXiv.2311.04279}

\bibitem[{Gordon {et~al.}(2003)Gordon, Clayton, Misselt, Landolt, \&
  Wolff}]{gordonQuantitativeComparisonSmall2003}
Gordon, K.~D., Clayton, G.~C., Misselt, K.~A., Landolt, A.~U., \& Wolff, M.~J.
  2003, The Astrophysical Journal, 594, 279, \dodoi{10.1086/376774}

\bibitem[{Grogin {et~al.}(2011)Grogin, Kocevski, Faber, Ferguson, Koekemoer,
  Riess, Acquaviva, Alexander, Almaini, Ashby, Barden, Bell, Bournaud, Brown,
  Caputi, Casertano, Cassata, Castellano, Challis, Chary, Cheung, Cirasuolo,
  Conselice, Roshan~Cooray, Croton, Daddi, Dahlen, Dav{\'e}, {de Mello}, Dekel,
  Dickinson, Dolch, Donley, Dunlop, Dutton, Elbaz, Fazio, Filippenko,
  Finkelstein, Fontana, Gardner, Garnavich, Gawiser, Giavalisco, Grazian, Guo,
  Hathi, H{\"a}ussler, Hopkins, Huang, Huang, Jha, Kartaltepe, Kirshner, Koo,
  Lai, Lee, Li, Lotz, Lucas, Madau, McCarthy, McGrath, McIntosh, McLure,
  Mobasher, Moustakas, Mozena, Nandra, Newman, Niemi, Noeske, Papovich,
  Pentericci, Pope, Primack, Rajan, Ravindranath, Reddy, Renzini, Rix, Robaina,
  Rodney, Rosario, Rosati, Salimbeni, Scarlata, Siana, Simard, Smidt,
  Somerville, Spinrad, Straughn, Strolger, Telford, Teplitz, Trump, {van der
  Wel}, Villforth, Wechsler, Weiner, Wiklind, Wild, Wilson, Wuyts, Yan, \&
  Yun}]{groginCANDELSCosmicAssembly2011}
Grogin, N.~A., Kocevski, D.~D., Faber, S.~M., {et~al.} 2011, The Astrophysical
  Journal Supplement Series, 197, 35, \dodoi{10.1088/0067-0049/197/2/35}

\bibitem[{Heintz {et~al.}(2025)Heintz, Brammer, Watson, Oesch, Keating, Hayes,
  {Abdurro'uf}, {Arellano-C{\'o}rdova}, Carnall, Christiansen, Cullen,
  Dav{\'e}, Dayal, Ferrara, Finlator, Fynbo, Flury, Gelli, Gillman,
  Gottumukkala, Gould, Greve, Hardin, Hsiao, Hutter, Jakobsson, Killi,
  Khosravaninezhad, Laursen, Lee, Magdis, Matthee, Naidu, Narayanan, Pollock,
  Prescott, Rusakov, Shuntov, Sneppen, Smit, Tanvir, Terp, Toft, Valentino,
  Vijayan, Weaver, Wise, \& Witstok}]{heintzJWSTPRIMALArchivalSurvey2025}
Heintz, K.~E., Brammer, G.~B., Watson, D., {et~al.} 2025, Astronomy and
  Astrophysics, 693, A60, \dodoi{10.1051/0004-6361/202450243}

\bibitem[{Johnson {et~al.}(2021)Johnson, Leja, Conroy, \&
  Speagle}]{johnsonStellarPopulationInference2021}
Johnson, B.~D., Leja, J., Conroy, C., \& Speagle, J.~S. 2021, The Astrophysical
  Journal Supplement Series, 254, 22, \dodoi{10.3847/1538-4365/abef67}

\bibitem[{Koekemoer {et~al.}(2011)Koekemoer, Faber, Ferguson, Grogin, Kocevski,
  Koo, Lai, Lotz, Lucas, McGrath, Ogaz, Rajan, Riess, Rodney, Strolger,
  Casertano, Castellano, Dahlen, Dickinson, Dolch, Fontana, Giavalisco,
  Grazian, Guo, Hathi, Huang, {van der Wel}, Yan, Acquaviva, Alexander,
  Almaini, Ashby, Barden, Bell, Bournaud, Brown, Caputi, Cassata, Challis,
  Chary, Cheung, Cirasuolo, Conselice, Roshan~Cooray, Croton, Daddi, Dav{\'e},
  {de Mello}, {de Ravel}, Dekel, Donley, Dunlop, Dutton, Elbaz, Fazio,
  Filippenko, Finkelstein, Frazer, Gardner, Garnavich, Gawiser, Gruetzbauch,
  Hartley, H{\"a}ussler, Herrington, Hopkins, Huang, Jha, Johnson, Kartaltepe,
  Khostovan, Kirshner, Lani, Lee, Li, Madau, McCarthy, McIntosh, McLure,
  McPartland, Mobasher, Moreira, Mortlock, Moustakas, Mozena, Nandra, Newman,
  Nielsen, Niemi, Noeske, Papovich, Pentericci, Pope, Primack, Ravindranath,
  Reddy, Renzini, Rix, Robaina, Rosario, Rosati, Salimbeni, Scarlata, Siana,
  Simard, Smidt, Snyder, Somerville, Spinrad, Straughn, Telford, Teplitz,
  Trump, Vargas, Villforth, Wagner, Wandro, Wechsler, Weiner, Wiklind, Wild,
  Wilson, Wuyts, \& Yun}]{koekemoerCANDELSCosmicAssembly2011}
Koekemoer, A.~M., Faber, S.~M., Ferguson, H.~C., {et~al.} 2011, The
  Astrophysical Journal Supplement Series, 197, 36,
  \dodoi{10.1088/0067-0049/197/2/36}

\bibitem[{Kriek \& Conroy(2013)}]{kriekDustAttenuationLaw2013}
Kriek, M., \& Conroy, C. 2013, The Astrophysical Journal, 775, L16,
  \dodoi{10.1088/2041-8205/775/1/L16}

\bibitem[{Kriek {et~al.}(2009)Kriek, {van Dokkum}, Labb{\'e}, Franx,
  Illingworth, Marchesini, \& Quadri}]{kriekUltraDeepNearInfraredSpectrum2009}
Kriek, M., {van Dokkum}, P.~G., Labb{\'e}, I., {et~al.} 2009, The Astrophysical
  Journal, 700, 221, \dodoi{10.1088/0004-637X/700/1/221}

\bibitem[{Lecroq {et~al.}(2025)Lecroq, Charlot, Bressan, Bruzual, Costa, Iorio,
  Mapelli, Santoliquido, Shepherd, \&
  Spera}]{lecroqNewPrescriptionSpectral2025}
Lecroq, M., Charlot, S., Bressan, A., {et~al.} 2025, Astronomy and
  Astrophysics, 695, A17, \dodoi{10.1051/0004-6361/202452463}

\bibitem[{Leitherer \&
  Heckman(1995)}]{leithererSyntheticPropertiesStarburst1995}
Leitherer, C., \& Heckman, T.~M. 1995, The Astrophysical Journal Supplement
  Series, 96, 9, \dodoi{10.1086/192112}

\bibitem[{Llerena {et~al.}(2025)Llerena, Pentericci, Napolitano, Mascia,
  Amor{\'i}n, Calabr{\`o}, Castellano, Cleri, Giavalisco, Grogin, Hathi,
  Hirschmann, Koekemoer, Nanayakkara, Pacucci, Shen, Wilkins, Yoon, Yung,
  Bhatawdekar, Lucas, Wang, Arrabal~Haro, Bagley, Finkelstein, Kartaltepe,
  Merlin, Papovich, Pirzkal, \& Santini}]{llerenaIonizingPhotonProduction2025}
Llerena, M., Pentericci, L., Napolitano, L., {et~al.} 2025, Astronomy and
  Astrophysics, 698, A302, \dodoi{10.1051/0004-6361/202453251}

\bibitem[{Luridiana {et~al.}(2015)Luridiana, Morisset, \&
  Shaw}]{luridianaPyNebNewTool2015}
Luridiana, V., Morisset, C., \& Shaw, R.~A. 2015, Astronomy and Astrophysics,
  573, A42, \dodoi{10.1051/0004-6361/201323152}

\bibitem[{Madau {et~al.}(2024)Madau, Giallongo, Grazian, \&
  Haardt}]{madauCosmicReionizationJWST2024}
Madau, P., Giallongo, E., Grazian, A., \& Haardt, F. 2024, The Astrophysical
  Journal, 971, 75, \dodoi{10.3847/1538-4357/ad5ce8}

\bibitem[{Maiolino {et~al.}(2024)Maiolino, Scholtz, {Curtis-Lake}, Carniani,
  Baker, {de Graaff}, Tacchella, {\"U}bler, D'Eugenio, Witstok, Curti, Arribas,
  Bunker, Charlot, Chevallard, Eisenstein, Egami, Ji, Jones, Lyu, Rawle,
  Robertson, Rujopakarn, Perna, Sun, Venturi, Williams, \&
  Willott}]{maiolinoJADESDiversePopulation2024}
Maiolino, R., Scholtz, J., {Curtis-Lake}, E., {et~al.} 2024, Astronomy and
  Astrophysics, 691, A145, \dodoi{10.1051/0004-6361/202347640}

\bibitem[{Maseda {et~al.}(2020)Maseda, Bacon, Lam, Matthee, Brinchmann, Schaye,
  Labbe, Schmidt, Boogaard, Bouwens, Cantalupo, Franx, Hashimoto, Inami,
  Kusakabe, Mahler, Nanayakkara, Richard, \&
  Wisotzki}]{masedaElevatedIonizingPhoton2020}
Maseda, M.~V., Bacon, R., Lam, D., {et~al.} 2020, Monthly Notices of the Royal
  Astronomical Society, 493, 5120, \dodoi{10.1093/mnras/staa622}

\bibitem[{Mason {et~al.}(2025)Mason, Chen, Stark, Lu, Topping, \&
  Tang}]{masonConstraints$zsim613$Intergalactic2025}
Mason, C.~A., Chen, Z., Stark, D.~P., {et~al.} 2025, Constraints on the
  \$z{\textbackslash}sim6-13\$ Intergalactic Medium from {{JWST}} Spectroscopy
  of {{Lyman-alpha}} Damping Wings in Galaxies,  arXiv,
  \dodoi{10.48550/arXiv.2501.11702}

\bibitem[{Mason {et~al.}(2019)Mason, Naidu, Tacchella, \&
  Leja}]{masonModelindependentConstraintsHydrogenionizing2019}
Mason, C.~A., Naidu, R.~P., Tacchella, S., \& Leja, J. 2019, Monthly Notices of
  the Royal Astronomical Society, 489, 2669, \dodoi{10.1093/mnras/stz2291}

\bibitem[{McGreer {et~al.}(2015)McGreer, Mesinger, \&
  D'Odorico}]{mcgreerModelindependentEvidenceFavour2015}
McGreer, I.~D., Mesinger, A., \& D'Odorico, V. 2015, Monthly Notices of the
  Royal Astronomical Society, 447, 499, \dodoi{10.1093/mnras/stu2449}

\bibitem[{Mu{\~n}oz {et~al.}(2024)Mu{\~n}oz, Mirocha, Chisholm, Furlanetto, \&
  Mason}]{munozReionizationJWSTPhoton2024a}
Mu{\~n}oz, J.~B., Mirocha, J., Chisholm, J., Furlanetto, S.~R., \& Mason, C.
  2024, Monthly Notices of the Royal Astronomical Society, 535, L37,
  \dodoi{10.1093/mnrasl/slae086}

\bibitem[{Nakane {et~al.}(2024)Nakane, Ouchi, Nakajima, Harikane, Ono, Umeda,
  Isobe, Zhang, \& Xu}]{nakaneLyaEmission132024}
Nakane, M., Ouchi, M., Nakajima, K., {et~al.} 2024, The Astrophysical Journal,
  967, 28, \dodoi{10.3847/1538-4357/ad38c2}

\bibitem[{Nanayakkara {et~al.}(2020)Nanayakkara, Brinchmann, Glazebrook,
  Bouwens, Kewley, Tran, Cowley, Fisher, Kacprzak, Labbe, \&
  Straatman}]{nanayakkaraReconstructingObservedIonizing2020}
Nanayakkara, T., Brinchmann, J., Glazebrook, K., {et~al.} 2020, The
  Astrophysical Journal, 889, 180, \dodoi{10.3847/1538-4357/ab65eb}

\bibitem[{Oesch {et~al.}(2023)Oesch, Brammer, Naidu, Bouwens, Chisholm,
  Illingworth, Matthee, Nelson, Qin, Reddy, Shapley, Shivaei, {van Dokkum},
  Weibel, Whitaker, Wuyts, {Covelo-Paz}, Endsley, Fudamoto, Giovinazzo,
  {Herard-Demanche}, Kerutt, Kramarenko, Labbe, Leonova, Lin, Magee,
  Marchesini, Maseda, Mason, Matharu, Meyer, Neufeld, Prieto~Lyon, Schaerer,
  Sharma, Shuntov, Smit, Stefanon, Wyithe, \& Xiao}]{oeschJWSTFRESCOSurvey2023}
Oesch, P.~A., Brammer, G., Naidu, R.~P., {et~al.} 2023, Monthly Notices of the
  Royal Astronomical Society, 525, 2864, \dodoi{10.1093/mnras/stad2411}

\bibitem[{Oke \& Gunn(1983)}]{okeSecondaryStandardStars1983}
Oke, J.~B., \& Gunn, J.~E. 1983, The Astrophysical Journal, 266, 713,
  \dodoi{10.1086/160817}

\bibitem[{Pacifici {et~al.}(2023)Pacifici, Iyer, Mobasher, {da Cunha},
  Acquaviva, Burgarella, Calistro~Rivera, Carnall, Chang, Chartab, Cooke,
  Fairhurst, Kartaltepe, Leja, Ma{\l}ek, Salmon, Torelli, {Vidal-Garc{\'i}a},
  Boquien, Brammer, Brown, Capak, Chevallard, Circosta, Croton, Davidzon,
  Dickinson, Duncan, Faber, Ferguson, Fontana, Guo, Haeussler, Hemmati,
  Jafariyazani, Kassin, Larson, Lee, Mantha, Marchi, Nayyeri, Newman, Pandya,
  Pforr, Reddy, Sanders, Shah, Shahidi, Stevans, Triani, Tyler, Vanderhoof, {de
  la Vega}, Wang, \& Weston}]{pacificiArtMeasuringPhysical2023}
Pacifici, C., Iyer, K.~G., Mobasher, B., {et~al.} 2023, The Astrophysical
  Journal, 944, 141, \dodoi{10.3847/1538-4357/acacff}

\bibitem[{Pahl {et~al.}(2025)Pahl, Topping, Shapley, Sanders, Reddy, Clarke,
  Kehoe, Bento, \& Brammer}]{pahlSpectroscopicAnalysisIonizing2025}
Pahl, A., Topping, M.~W., Shapley, A., {et~al.} 2025, The Astrophysical
  Journal, 981, 134, \dodoi{10.3847/1538-4357/adb1ab}

\bibitem[{Pahl {et~al.}(2021)Pahl, Shapley, Steidel, Chen, \&
  Reddy}]{pahlUncontaminatedMeasurementEscaping2021}
Pahl, A.~J., Shapley, A., Steidel, C.~C., Chen, Y., \& Reddy, N.~A. 2021,
  Monthly Notices of the Royal Astronomical Society, 505, 2447,
  \dodoi{10.1093/mnras/stab1374}

\bibitem[{Papovich {et~al.}(2025)Papovich, Cole, Hu, Finkelstein, Shen,
  Arrabal~Haro, Amor{\'i}n, Backhaus, Bagley, Bhatawdekar, Calabr{\'o},
  Carnall, Cleri, Daddi, Dickinson, Grogin, Holwerda, Jaskot, Koekemoer,
  Llerena, Lucas, Mascia, Pacucci, Pentericci, {P{\'e}rez-Gonz{\'a}lez},
  Pirzkal, Raghunathan, Seill{\'e}, Somerville, \&
  Yung}]{papovichGalaxiesEpochReionization2025}
Papovich, C., Cole, J.~W., Hu, W., {et~al.} 2025, Galaxies in the {{Epoch}} of
  {{Reionization Are All Bark}} and {{No Bite}} -- {{Plenty}} of {{Ionizing
  Photons}}, {{Low Escape Fractions}},  arXiv,
  \dodoi{10.48550/arXiv.2505.08870}

\bibitem[{{Planck Collaboration} {et~al.}(2016){Planck Collaboration}, Ade,
  Aghanim, Arnaud, Ashdown, Aumont, Baccigalupi, Banday, Barreiro, Bartlett,
  Bartolo, Battaner, Battye, Benabed, Beno{\^i}t, {Benoit-L{\'e}vy}, Bernard,
  Bersanelli, Bielewicz, Bock, Bonaldi, Bonavera, Bond, Borrill, Bouchet,
  Boulanger, Bucher, Burigana, Butler, Calabrese, Cardoso, Catalano, Challinor,
  Chamballu, Chary, Chiang, Chluba, Christensen, Church, Clements, Colombi,
  Colombo, Combet, Coulais, Crill, Curto, Cuttaia, Danese, Davies, Davis,
  De~Bernardis, De~Rosa, De~Zotti, Delabrouille, D{\'e}sert, Di~Valentino,
  Dickinson, Diego, Dolag, Dole, Donzelli, Dor{\'e}, Douspis, Ducout, Dunkley,
  Dupac, Efstathiou, Elsner, En{\ss}lin, Eriksen, Farhang, Fergusson, Finelli,
  Forni, Frailis, Fraisse, Franceschi, Frejsel, Galeotta, Galli, Ganga,
  Gauthier, Gerbino, Ghosh, Giard, {Giraud-H{\'e}raud}, Giusarma, Gjerl{\o}w,
  {Gonz{\'a}lez-Nuevo}, G{\'o}rski, Gratton, Gregorio, Gruppuso, Gudmundsson,
  Hamann, Hansen, Hanson, Harrison, Helou, {Henrot-Versill{\'e}},
  {Hern{\'a}ndez-Monteagudo}, Herranz, Hildebrandt, Hivon, Hobson, Holmes,
  Hornstrup, Hovest, Huang, Huffenberger, Hurier, Jaffe, Jaffe, Jones, Juvela,
  Keih{\"a}nen, Keskitalo, Kisner, Kneissl, Knoche, Knox, Kunz, {Kurki-Suonio},
  Lagache, L{\"a}hteenm{\"a}ki, Lamarre, Lasenby, Lattanzi, Lawrence, Leahy,
  Leonardi, Lesgourgues, Levrier, Lewis, Liguori, Lilje, {Linden-V{\o}rnle},
  {L{\'o}pez-Caniego}, Lubin, {Maci{\'a}s-P{\'e}rez}, Maggio, Maino, Mandolesi,
  Mangilli, Marchini, Maris, Martin, Martinelli, {Mart{\'i}nez-Gonz{\'a}lez},
  Masi, Matarrese, Mcgehee, Meinhold, Melchiorri, Melin, Mendes, Mennella,
  Migliaccio, Millea, Mitra, {Miville-Desch{\^e}nes}, Moneti, Montier,
  Morgante, Mortlock, Moss, Munshi, Murphy, Naselsky, Nati, Natoli,
  Netterfield, {N{\o}rgaard-Nielsen}, Noviello, Novikov, Novikov, Oxborrow,
  Paci, Pagano, Pajot, Paladini, Paoletti, Partridge, Pasian, Patanchon,
  Pearson, Perdereau, Perotto, Perrotta, Pettorino, Piacentini, Piat,
  Pierpaoli, Pietrobon, Plaszczynski, Pointecouteau, Polenta, Popa, Pratt,
  Pr{\'e}zeau, Prunet, Puget, Rachen, Reach, Rebolo, Reinecke, Remazeilles,
  Renault, Renzi, Ristorcelli, Rocha, Rosset, Rossetti, Roudier,
  Rouill{\'e}~D'orfeuil, {Rowan-Robinson}, {Rubin{\~o}-Mart{\'i}n}, Rusholme,
  Said, Salvatelli, Salvati, Sandri, Santos, Savelainen, Savini, Scott,
  Seiffert, Serra, Shellard, Spencer, Spinelli, Stolyarov, Stompor, Sudiwala,
  Sunyaev, Sutton, {Suur-Uski}, Sygnet, Tauber, Terenzi, Toffolatti, Tomasi,
  Tristram, Trombetti, Tucci, Tuovinen, T{\"u}rler, Umana, Valenziano,
  Valiviita, Van~Tent, Vielva, Villa, Wade, Wandelt, Wehus, White, White,
  Wilkinson, Yvon, Zacchei, \&
  Zonca}]{planckcollaborationPlanck2015Results2016}
{Planck Collaboration}, Ade, P.~A., Aghanim, N., {et~al.} 2016, Astronomy and
  Astrophysics, 594, \dodoi{10.1051/0004-6361/201525830}

\bibitem[{Price {et~al.}(2014)Price, Kriek, Brammer, Conroy,
  F{\"o}rster~Schreiber, Franx, Fumagalli, Lundgren, Momcheva, Nelson, Skelton,
  {van Dokkum}, Whitaker, \& Wuyts}]{priceDirectMeasurementsDust2014}
Price, S.~H., Kriek, M., Brammer, G.~B., {et~al.} 2014, The Astrophysical
  Journal, 788, 86, \dodoi{10.1088/0004-637X/788/1/86}

\bibitem[{{Prieto-Lyon} {et~al.}(2023){Prieto-Lyon}, Strait, Mason, Brammer,
  Caminha, Mercurio, Acebron, Bergamini, Grillo, Rosati, Vanzella, Castellano,
  Merlin, Paris, Boyett, Calabr{\`o}, Morishita, Mascia, Pentericci,
  {Roberts-Borsani}, Roy, Treu, \&
  Vulcani}]{prieto-lyonProductionIonizingPhotons2023}
{Prieto-Lyon}, G., Strait, V., Mason, C.~A., {et~al.} 2023, Astronomy and
  Astrophysics, 672, A186, \dodoi{10.1051/0004-6361/202245532}

\bibitem[{Puglisi {et~al.}(2016)Puglisi, Rodighiero, Franceschini, Talia,
  Cimatti, Baronchelli, Daddi, Renzini, Schawinski, Mancini, Silverman,
  Gruppioni, Lutz, Berta, \& Oliver}]{puglisiDustAttenuation12016}
Puglisi, A., Rodighiero, G., Franceschini, A., {et~al.} 2016, Astronomy and
  Astrophysics, 586, A83, \dodoi{10.1051/0004-6361/201526782}

\bibitem[{Reddy {et~al.}(2023)Reddy, Topping, Sanders, Shapley, \&
  Brammer}]{reddyPaschenlineConstraintsDust2023a}
Reddy, N.~A., Topping, M.~W., Sanders, R.~L., Shapley, A.~E., \& Brammer, G.
  2023, ApJ, 948, 83, \dodoi{10.3847/1538-4357/acc869}

\bibitem[{Reddy {et~al.}(2015)Reddy, Kriek, Shapley, Freeman, Siana, Coil,
  Mobasher, Price, Sanders, \& Shivaei}]{reddyMOSDEFSURVEYMEASUREMENTS2015}
Reddy, N.~A., Kriek, M., Shapley, A.~E., {et~al.} 2015, The Astrophysical
  Journal, 806, 259, \dodoi{10.1088/0004-637X/806/2/259}

\bibitem[{Reddy {et~al.}(2018)Reddy, Oesch, Bouwens, Montes, Illingworth,
  Steidel, {van Dokkum}, Atek, Carollo, Cibinel, Holden, Labb{\'e}, Magee,
  Morselli, Nelson, \& Wilkins}]{reddyHDUVSurveyRevised2018}
Reddy, N.~A., Oesch, P.~A., Bouwens, R.~J., {et~al.} 2018, The Astrophysical
  Journal, 853, 56, \dodoi{10.3847/1538-4357/aaa3e7}

\bibitem[{Reddy {et~al.}(2020)Reddy, Shapley, Kriek, Steidel, Shivaei, Sanders,
  Mobasher, Coil, Siana, Freeman, Azadi, Fetherolf, Leung, Price, \&
  Zick}]{reddyMOSDEFSurveyFirst2020}
Reddy, N.~A., Shapley, A.~E., Kriek, M., {et~al.} 2020, The Astrophysical
  Journal, 902, 123, \dodoi{10.3847/1538-4357/abb674}

\bibitem[{Reddy {et~al.}(2025)Reddy, Shapley, Sanders, Topping, Ellis, Pettini,
  Brammer, Cullen, Forster~Schreiber, Khostovan, McLeod, McLure, Narayanan,
  Oesch, Pahl, Steidel, \& Berg}]{reddyJWSTAURORASurvey2025}
Reddy, N.~A., Shapley, A.~E., Sanders, R.~L., {et~al.} 2025, The
  {{JWST}}/{{AURORA Survey}}: {{Multiple Balmer}} and {{Paschen Emission
  Lines}} for {{Individual Star-forming Galaxies}} at Z=1.5-4.4. {{I}}. {{A
  Diversity}} of {{Nebular Attenuation Curves}} and {{Evidence}} for
  {{Non-Unity Dust Covering Fractions}},  arXiv,
  \dodoi{10.48550/arXiv.2506.17396}

\bibitem[{Robertson {et~al.}(2015)Robertson, Ellis, Furlanetto, \&
  Dunlop}]{robertsonCosmicReionizationEarly2015}
Robertson, B.~E., Ellis, R.~S., Furlanetto, S.~R., \& Dunlop, J.~S. 2015, The
  Astrophysical Journal Letters, 802, L19, \dodoi{10.1088/2041-8205/802/2/L19}

\bibitem[{Ronayne {et~al.}(2024)Ronayne, Papovich, Yang, Shen, Dickinson,
  Kennicutt, Alavi, Arrabal~Haro, Bagley, Burgarella, Le~Bail, Bell, Cleri,
  Cole, Costantin, {de la Vega}, Daddi, Elbaz, Finkelstein, Grogin, Holwerda,
  Kartaltepe, Kirkpatrick, Koekemoer, Lucas, Magnelli, Mobasher,
  {P{\'e}rez-Gonz{\'a}lez}, Prichard, Rafelski, Rodighiero, Sunnquist, Teplitz,
  Wang, Windhorst, \& Yung}]{ronayneCEERS77Mm2024}
Ronayne, K., Papovich, C., Yang, G., {et~al.} 2024, The Astrophysical Journal,
  970, 61, \dodoi{10.3847/1538-4357/ad5006}

\bibitem[{Rosdahl {et~al.}(2018)Rosdahl, Katz, Blaizot, Kimm, {Michel-Dansac},
  Garel, Haehnelt, Ocvirk, \&
  Teyssier}]{rosdahlSPHINXCosmologicalSimulations2018}
Rosdahl, J., Katz, H., Blaizot, J., {et~al.} 2018, Monthly Notices of the Royal
  Astronomical Society, 479, 994, \dodoi{10.1093/mnras/sty1655}

\bibitem[{Salim {et~al.}(2018)Salim, Boquien, \&
  Lee}]{salimDustAttenuationCurves2018}
Salim, S., Boquien, M., \& Lee, J.~C. 2018, The Astrophysical Journal, 859, 11,
  \dodoi{10.3847/1538-4357/aabf3c}

\bibitem[{Sanders {et~al.}(2024)Sanders, Shapley, Topping, Reddy, \&
  Brammer}]{sandersDirectEbasedMetallicities2024}
Sanders, R.~L., Shapley, A.~E., Topping, M.~W., Reddy, N.~A., \& Brammer, G.~B.
  2024, The Astrophysical Journal, 962, 24, \dodoi{10.3847/1538-4357/ad15fc}

\bibitem[{Sanders {et~al.}(2025{\natexlab{a}})Sanders, Shapley, Topping, Reddy,
  Berg, Bouwens, Brammer, Carnall, Cullen, Dav{\'e}, Dunlop, Ellis,
  F{\"o}rster~Schreiber, Furlanetto, Glazebrook, Illingworth, Jones, Kriek,
  McLeod, McLure, Narayanan, Oesch, Pahl, Pettini, Schaerer, Stark, Steidel,
  Tang, Clarke, Donnan, \& Kehoe}]{sandersAURORASurveyNebular2025}
Sanders, R.~L., Shapley, A.~E., Topping, M.~W., {et~al.} 2025{\natexlab{a}},
  The Astrophysical Journal, 989, 209, \dodoi{10.3847/1538-4357/adf066}

\bibitem[{Sanders {et~al.}(2025{\natexlab{b}})Sanders, Shapley, Topping, Reddy,
  Berg, Khostovan, Bouwens, Brammer, Carnall, Cullen, Dav{\'e}, Dunlop, Ellis,
  F{\"o}rster~Schreiber, Furlanetto, Glazebrook, Illingworth, Jones, Kriek,
  McLeod, McLure, Narayanan, Oesch, Pahl, Pettini, Schaerer, Stark, Steidel,
  Tang, Clarke, Donnan, \& Kehoe}]{sandersAURORASurveyHighRedshift2025}
---. 2025{\natexlab{b}}, The {{AURORA Survey}}: {{High-Redshift Empirical
  Metallicity Calibrations}} from {{Electron Temperature Measurements}} at
  Z=2-10,  arXiv, \dodoi{10.48550/arXiv.2508.10099}

\bibitem[{Saxena {et~al.}(2024)Saxena, Bunker, Jones, Stark, Cameron, Witstok,
  Arribas, Baker, Baum, Bhatawdekar, Bowler, Boyett, Carniani, Charlot,
  Chevallard, Curti, {Curtis-Lake}, Eisenstein, Endsley, Hainline, Helton,
  Johnson, Kumari, Looser, Maiolino, Rieke, Rix, Robertson, Sandles, Simmonds,
  Smit, Tacchella, Williams, Willmer, \&
  Willott}]{saxenaJADESProductionEscape2024}
Saxena, A., Bunker, J.~A., Jones, C.~G., {et~al.} 2024, A\&A, 684, A84,
  \dodoi{10.1051/0004-6361/202347132}

\bibitem[{Shapley {et~al.}(2023)Shapley, Sanders, Reddy, Topping, \&
  Brammer}]{shapleyJWSTNIRSpecBalmerline2023a}
Shapley, A.~E., Sanders, R.~L., Reddy, N.~A., Topping, M.~W., \& Brammer, G.~B.
  2023, The Astrophysical Journal, 954, 157, \dodoi{10.3847/1538-4357/acea5a}

\bibitem[{Shapley {et~al.}(2025)Shapley, Sanders, Topping, Reddy, Berg,
  Bouwens, Brammer, Carnall, Cullen, Dav{\'e}, Dunlop, Ellis,
  F{\"o}rster~Schreiber, Furlanetto, Glazebrook, Illingworth, Jones, Kriek,
  McLeod, McLure, Narayanan, Oesch, Pahl, Pettini, Schaerer, Stark, Steidel,
  Tang, Clarke, Donnan, \& Kehoe}]{shapleyAURORASurveyNew2025}
Shapley, A.~E., Sanders, R.~L., Topping, M.~W., {et~al.} 2025, The
  Astrophysical Journal, 980, 242, \dodoi{10.3847/1538-4357/adad68}

\bibitem[{Shen {et~al.}(2020)Shen, Hopkins, {Faucher-Gigu{\`e}re}, Alexander,
  Richards, Ross, \& Hickox}]{shenBolometricQuasarLuminosity2020}
Shen, X., Hopkins, P.~F., {Faucher-Gigu{\`e}re}, C.~A., {et~al.} 2020, Monthly
  Notices of the Royal Astronomical Society, 495, 3252,
  \dodoi{10.1093/mnras/staa1381}

\bibitem[{Shivaei \& Boogaard(2024)}]{shivaeiTightCorrelationPAH2024}
Shivaei, I., \& Boogaard, L. 2024, The Tight Correlation of {{PAH}} and {{CO}}
  Emission from Z{\textasciitilde}0-4, \dodoi{10.48550/arXiv.2409.05710}

\bibitem[{Shivaei {et~al.}(2018)Shivaei, Reddy, Siana, Shapley, Kriek,
  Mobasher, Freeman, Sanders, Coil, Price, Fetherolf, Azadi, Leung, \&
  Zick}]{shivaeiMOSDEFSurveyDirect2018}
Shivaei, I., Reddy, N.~A., Siana, B., {et~al.} 2018, The Astrophysical Journal,
  855, 42, \dodoi{10.3847/1538-4357/aaad62}

\bibitem[{Simmonds {et~al.}(2024{\natexlab{a}})Simmonds, Tacchella, Hainline,
  Johnson, Pusk{\'a}s, Robertson, Baker, Bhatawdekar, Boyett, Bunker, Cargile,
  Carniani, Chevallard, Curti, {Curtis-Lake}, Ji, Jones, Kumari, Laseter,
  Maiolino, Maseda, Rinaldi, Stoffers, {\"U}bler, Villanueva, Williams,
  Willott, Witstok, \& Zhu}]{simmondsIonizingPropertiesGalaxies2024}
Simmonds, C., Tacchella, S., Hainline, K., {et~al.} 2024{\natexlab{a}}, Monthly
  Notices of the Royal Astronomical Society, 535, 2998,
  \dodoi{10.1093/mnras/stae2537}

\bibitem[{Simmonds {et~al.}(2024{\natexlab{b}})Simmonds, Tacchella, Hainline,
  Johnson, McClymont, Robertson, Saxena, Sun, Witten, Baker, Bhatawdekar,
  Boyett, Bunker, Charlot, {Curtis-Lake}, Egami, Eisenstein, Hausen, Maiolino,
  Maseda, Scholtz, Williams, Willott, \&
  Witstok}]{simmondsLowmassBurstyGalaxies2024}
---. 2024{\natexlab{b}}, Monthly Notices of the Royal Astronomical Society,
  527, 6139, \dodoi{10.1093/mnras/stad3605}

\bibitem[{Skelton {et~al.}(2014)Skelton, Whitaker, Momcheva, Brammer,
  Van~Dokkum, Labb{\'e}, Franx, Van Der~Wel, Bezanson, Da~Cunha, Fumagalli,
  F{\"o}rster~Schreiber, Kriek, Leja, Lundgren, Magee, Marchesini, Maseda,
  Nelson, Oesch, Pacifici, Patel, Price, Rix, Tal, Wake, \&
  Wuyts}]{skelton3DHSTWFC3selectedPhotometric2014}
Skelton, R.~E., Whitaker, K.~E., Momcheva, I.~G., {et~al.} 2014, Astrophysical
  Journal, Supplement Series, 214, 24, \dodoi{10.1088/0067-0049/214/2/24}

\bibitem[{Smith {et~al.}(2007)Smith, Draine, Dale, Moustakas, Kennicutt, Helou,
  Armus, Roussel, Sheth, Bendo, Buckalew, Calzetti, Engelbracht, Gordon,
  Hollenbach, Li, Malhotra, Murphy, \&
  Walter}]{smithMidInfraredSpectrumStarforming2007}
Smith, J. D.~T., Draine, B.~T., Dale, D.~A., {et~al.} 2007, The Astrophysical
  Journal, 656, 770, \dodoi{10.1086/510549}

\bibitem[{Stanway {et~al.}(2016)Stanway, Eldridge, \&
  Becker}]{stanwayStellarPopulationEffects2016}
Stanway, E.~R., Eldridge, J.~J., \& Becker, G.~D. 2016, Monthly Notices of the
  Royal Astronomical Society, 456, 485, \dodoi{10.1093/mnras/stv2661}

\bibitem[{Steidel {et~al.}(2018)Steidel, Bogosavljevi{\'c}, Shapley, Reddy,
  Rudie, Pettini, Trainor, \& Strom}]{steidelKeckLymanContinuum2018}
Steidel, C.~C., Bogosavljevi{\'c}, M., Shapley, A.~E., {et~al.} 2018, The
  Astrophysical Journal, 869, 123, \dodoi{10.3847/1538-4357/aaed28}

\bibitem[{Tang {et~al.}(2019)Tang, Stark, Chevallard, \&
  Charlot}]{tangMMTMMIRSSpectroscopy2019}
Tang, M., Stark, D.~P., Chevallard, J., \& Charlot, S. 2019, Monthly Notices of
  the Royal Astronomical Society, 489, 2572, \dodoi{10.1093/mnras/stz2236}

\bibitem[{Topping {et~al.}(2022)Topping, Stark, Endsley, Plat, Whitler, Chen,
  \& Charlot}]{toppingSearchingExtremelyBlue2022}
Topping, M.~W., Stark, D.~P., Endsley, R., {et~al.} 2022, The Astrophysical
  Journal, 941, 153, \dodoi{10.3847/1538-4357/aca522}

\bibitem[{Topping {et~al.}(2025)Topping, Sanders, Shapley, Pahl, Reddy, Stark,
  Berg, Clarke, Cullen, Dunlop, Ellis, F{\"o}rster~Schreiber, Illingworth,
  Jones, Narayanan, Pettini, \& Schaerer}]{toppingAURORASurveyEvolution2025}
Topping, M.~W., Sanders, R.~L., Shapley, A.~E., {et~al.} 2025, The {{AURORA
  Survey}}: {{The Evolution}} of {{Multi-phase Electron Densities}} at {{High
  Redshift}},  arXiv, \dodoi{10.48550/arXiv.2502.08712}

\bibitem[{Treu {et~al.}(2022)Treu, {Roberts-Borsani}, Bradac, Brammer, Fontana,
  Henry, Mason, Morishita, Pentericci, Wang, Acebron, Bagley, Bergamini,
  Belfiori, Bonchi, Boyett, Boutsia, Calabr{\'o}, Caminha, Castellano,
  Dressler, Glazebrook, Grillo, Jacobs, Jones, Kelly, Leethochawalit, Malkan,
  Marchesini, Mascia, Mercurio, Merlin, Nanayakkara, Nonino, Paris, Poggianti,
  Rosati, Santini, Scarlata, Shipley, Strait, Trenti, Tubthong, Vanzella,
  Vulcani, \& Yang}]{treuGLASSJWSTEarlyRelease2022}
Treu, T., {Roberts-Borsani}, G., Bradac, M., {et~al.} 2022, The Astrophysical
  Journal, 935, 110, \dodoi{10.3847/1538-4357/ac8158}

\bibitem[{Valentino {et~al.}(2023)Valentino, Brammer, Gould, Kokorev, Fujimoto,
  Jespersen, Vijayan, Weaver, Ito, Tanaka, Ilbert, Magdis, Whitaker, Faisst,
  Gallazzi, Gillman, {Gim{\'e}nez-Arteaga}, {G{\'o}mez-Guijarro}, Kubo, Heintz,
  Hirschmann, Oesch, Onodera, Rizzo, Lee, Strait, \&
  Toft}]{valentinoAtlasColorselectedQuiescent2023}
Valentino, F., Brammer, G., Gould, K. M.~L., {et~al.} 2023, The Astrophysical
  Journal, 947, 20, \dodoi{10.3847/1538-4357/acbefa}

\bibitem[{Vanzella {et~al.}(2012)Vanzella, Guo, Giavalisco, Grazian,
  Castellano, Cristiani, Dickinson, Fontana, Nonino, Giallongo, Pentericci,
  Galametz, Faber, Ferguson, Grogin, Koekemoer, Newman, \&
  Siana}]{vanzellaDetectionIonizingRadiation2012}
Vanzella, E., Guo, Y., Giavalisco, M., {et~al.} 2012, The Astrophysical
  Journal, 751, 70, \dodoi{10.1088/0004-637X/751/1/70}

\bibitem[{Wild {et~al.}(2011)Wild, Charlot, Brinchmann, Heckman, Vince,
  Pacifici, \& Chevallard}]{wildEmpiricalDeterminationShape2011}
Wild, V., Charlot, S., Brinchmann, J., {et~al.} 2011, Monthly Notices of the
  Royal Astronomical Society, 417, 1760,
  \dodoi{10.1111/j.1365-2966.2011.19367.x}

\bibitem[{Williams {et~al.}(2023)Williams, Tacchella, Maseda, Robertson,
  Johnson, Willott, Eisenstein, Willmer, Ji, Hainline, Helton, Alberts, Baum,
  Bhatawdekar, Boyett, Bunker, Carniani, Charlot, Chevallard, {Curtis-Lake},
  {de Graaff}, Egami, Franx, Kumari, Maiolino, Nelson, Rieke, Sandles, Shivaei,
  Simmonds, Smit, Suess, Sun, {\"U}bler, \&
  Witstok}]{williamsJEMSDeepMediumband2023}
Williams, C.~C., Tacchella, S., Maseda, M.~V., {et~al.} 2023, The Astrophysical
  Journal Supplement Series, 268, 64, \dodoi{10.3847/1538-4365/acf130}

\end{thebibliography}

	\end{document}